\documentclass[a4paper,11pt]{article}
\pdfoutput=1 

\usepackage{jcappub} 
\usepackage{natbib}
\usepackage[T1]{fontenc} 
\usepackage{verbatim}
\usepackage{color}

\newcommand{\Ggpl}{\Gamma_{\gamma \mathrm{Pl}}}
\newcommand{\GgX}{\Gamma_{\gamma X}}
\newcommand{\cm}{\mathrm{cm}}
\newcommand{\cmsq}{\mathrm{cm}^2}
\newcommand{\gram}{\mathrm{g}}
\newcommand{\eV}{\mathrm{eV}}
\newcommand{\MeV}{\mathrm{MeV}}

\newcommand{\sx}{\sigma_X}
\newcommand{\Mx}{M_X}
\newcommand{\sM}{\sx/\Mx}

\newcommand{\zin}{z_\mathrm{in}}

\title{Updated cosmological constraints on Macroscopic Dark Matter}

\author[a,b]{Luca Caloni}
\author[b]{Martina Gerbino}
\author[b]{Massimiliano Lattanzi}
\affiliation[a]{Dipartimento di Fisica e Scienze della Terra, Universit\`a degli Studi di Ferrara, via Giuseppe Saragat 1, I-44122 Ferrara, Italy}
\affiliation[b]{Istituto Nazionale di Fisica Nucleare, Sezione di Ferrara, via Giuseppe Saragat 1, I-44122 Ferrara, Italy}

\emailAdd{luca.caloni@unife.it}
\emailAdd{gerbinom@fe.infn.it}
\emailAdd{lattanzi@fe.infn.it}

\abstract{We revise the cosmological phenomenology of Macroscopic Dark Matter (MDM) candidates, also commonly dubbed as Macros. A possible signature of MDM is the capture of baryons from the cosmological plasma in the pre-recombination epoch, with the consequent injection of high-energy photons in the baryon-photon plasma.  
By keeping a phenomenological approach, we consider two broad classes of MDM in which Macros are composed either of ordinary matter or antimatter. In both scenarios, we also analyze the impact of a non-vanishing electric charge carried by Macros. 
We derive constraints on the Macro parameter space from three cosmological processes: the change in the baryon density between the end of the Big Bang Nucleosynthesis (BBN) and the Cosmic Microwave Background (CMB) decoupling, the production of spectral distortions in the CMB and the kinetic coupling between charged MDM and baryons at the time of recombination. 
In the case of neutral Macros we find that the tightest constraints are set by the baryon density condition in most of the parameter space. For Macros composed of ordinary matter and with binding energy $I$, this leads to the following bound on the reduced cross-section: $\sigma_X/M_X \lesssim 6.8 \cdot 10^{-7} \left(I/\mathrm{MeV}\right)^{-1.56} \, \text{cm}^2 \, \text{g}^{-1}$. Charged Macros with surface potential $V_X$, instead, are mainly constrained by the tight coupling with baryons, resulting in $\sigma_X/M_X \lesssim 2 \cdot 10^{-11} \left(|V_X|/\mathrm{MeV}\right)^{-2} \text{cm}^2 \, \text{g}^{-1}$. 
Finally, we show that future CMB spectral distortions experiments, like PIXIE and SuperPIXIE, would have the sensitivity to probe larger regions of the parameter space: this would allow either for a possible evidence or for an improvement of the current bounds on Macros as dark matter candidates.}

\begin{document}
\maketitle
\flushbottom

\section{Introduction}

Understanding the nature of dark matter is one of the major endeavours of modern cosmology and particle physics. Over the years, a plethora of models have been proposed, most of which explain the dark matter as a new particle arising in some extension of the Standard Model of particle physics. However, given the lack of experimental evidence in laboratory searches, it is important to keep an open mind on alternative scenarios, some of which could be realized within the Standard Model itself. An appealing possibility is that dark matter consists of macroscopic-size (in a sense to be specified later) objects, generically dubbed ``macro'' dark matter (MDM). Over the years, many models that fall into this category have been proposed, such as Strangelets \cite{Witten:1984rs,Alcock:1985vc,Alcock:1986hz,Farhi:1984qu,LYNN1990186}, Q-balls \cite{Coleman:1985ki,Lee:1991ax,Kusenko:1997si,Kusenko:2004yw,Kusenko:1997vp} and Compact Composite Objects \cite{Zhitnitsky:2006vt,Zhitnitsky:2006tu,Cumberbatch:2006bj,Bai:2020jfm,Cline:2013zca}. Primordial black holes, that have gained much attention following the LIGO and VIRGO detection of the gravitational wave signal from black hole mergers, are also an example of MDM\footnote{Constraints on PBHs are usually the weakest that can be obtained for a given MDM mass. In fact, a black hole is the smallest object that can exist with a given mass, and this implies it also has the smallest possible geometric cross section.} (see e.g. \cite{Carr:2016drx,doi:10.1146/annurev-nucl-050520-125911} for a review).  A broad class of MDM candidates includes yet-unknown, composite states of baryonic matter. The prototype of this kind of MDM are strangelets, consisting of up, down and strange quarks confined in a quark phase \cite{Witten:1984rs,Alcock:1985vc,Alcock:1986hz,Farhi:1984qu,LYNN1990186}. 
Interestingly enough, the existence of such a phase of quark matter might help explaining the nature of compact objects originating
the gravitational wave event GW190814 observed by LIGO and VIRGO \cite{Abbott:2020khf,Bombaci:2020vgw}. 
Scenarios in which the DM is still a composite object, but based on physics beyond the Standard Model have also been envisaged, see e.g. \cite{Zhitnitsky:2002nr,Zhitnitsky:2004da,Liang:2016tqc,Bai:2018dxf,Bai:2018vik}.
A remarkable feature is that many of these models allow us to account for the dark matter within the Standard Model of particle physics, i.e. without requiring the introduction of additional particle species. In particular, in this paper we focus on the possibility that Macros are composed of free quarks or anti-quarks. 

In this paper we examine the cosmological phenomenology of MDM. We choose not to make explicit reference to any specific model. However, we focus on a particular process that might be associated to MDM, namely proton capture by dark matter. A proposal to look for signatures of this process in large-volume terrestrial detectors (like e.g., neutrino detectors) has been advanced in Ref. \cite{Bai:2019ogh}.
Cosmological imprints of MDM have been previously examined in the literature, see e.g. Refs. \cite{Jacobs:2014yca,Sidhu:2019gwo,SinghSidhu:2020cxw,Bai:2020jfm,Kumar:2018rlf} (for other constraints on MDM, see also \cite{Bai:2019ogh,Sidhu:2019qoa,Sidhu:2019fgg,Sidhu:2019oii,Brandt:2016aco,Sidhu:2019kpd,Starkman:2020sbz,Zumalacarregui:2017qqd,Kuhnel:2017bvu}). In comparison to previous works, we examine more carefully the peculiar signatures of proton capture, starting from considerations related to the chemical equilibrium of the relevant reactions. In more detail, from a phenomenological standpoint, we analyze two ``benchmark'' scenarios. In the first, proton capture results in the Macro $X$ transitioning to a more stable state $X'$ with energy release in the plasma. The amount of energy released is set by the binding energy of the MDM, which we treat as a free parameter. Conversely, the newly formed $X'$ can be photodissociated by background photons. The time at which the direct and inverse processes go out of equilibrium is determined by the binding energy of Macros, which then fixes the moment in the cosmological evolution when the comoving density of baryons starts decreasing. This first scenario can be thought as representative of models of ``nuclear'' MDM (e.g. strangelets). In the second scenario, we consider Macros that contain antibaryons. In this case, proton capture results in annihilation processes, similarly to what happens with proton capture over ordinary antinuclei (see e.g. \cite{Richard:2019dic}). This also results in energy injection in the primordial plasma; however, differently from the first scenario, there is no inverse process by which the starting Macro population can be replenished. In both cases, we also account for the possibility that Macros carry a non-vanishing electric charge.

We consider absorption of protons happening after the end of Big Bang Nucleosynthesis (BBN)\footnote{In the case of Macros composed of ordinary matter, this translates into a requirement on the values of the binding energy that we consider.}. We focus on the following probes of proton capture: change of baryon density between the end of BBN and the cosmic microwave background (CMB) decoupling; spectral distortions in the CMB; kinetic coupling between MDM and baryons around the time of hydrogen recombination. We also assess the impact on light element abundances, that might be changed after the end of BBN due to the energy injection associated to proton capture. We provide constraints from current observations, and discuss detection prospects from future experiments.

The paper is organized as follows. In Section \ref{sec:Macros} we focus on the scenario of Macros composed of ordinary matter. We first make some thermodynamical considerations on the reactions involved in the process of proton capture and the consequent photon emission. Then, we derive some constraints on the Macro parameter space from the three cosmological probes mentioned above. In Section \ref{sec:anti-Macros} we repeat a similar analysis for the case of Macros composed of antimatter. In Section \ref{sec:Conclusions} we discuss our results, also in light of the constraints derived in previous literature, and draw our conclusions.

\section{Macroscopic Dark Matter}
\label{sec:Macros}
  
In this section of the paper we consider the case of Macros composed of ordinary matter. As stated in the introduction, the picture we have in mind is that of Macros composed of free quarks. After discussing about the chemical equilibrium of the relevant interactions at play, we derive constraints on the reduced cross-section of Macros, $\sigma_X/M_X$, by analyzing the cosmological phenomenology of MDM.

\subsection{Thermodynamics of Macro DM}
\label{subsec:Thermodynamics}

We consider the following process, in which the dark matter $X$ captures a baryon (a neutron or proton), emitting a photon:
\begin{equation}
X^N + B \longleftrightarrow X^{N+1} + \gamma \, . \label{eq:bcapture}
\end{equation}
Here, $N$ is the baryon number of the MDM particle. We assume that the absorption process can continue indefinitely, i.e. that $X^{N+1}$ can absorb another baryon to give $X^{N+2}$, emitting a photon of (roughly) the same energy, and so on. In principle, the inverse process is also possible: a photon with enough energy can rip a proton or neutron off the DM. 

One thing to stress is that, if the emitted photon interacts with another MDM particle on timescales short with respect to the expansion rate,    the numbers of $X^N$'s, protons and neutrons are separately conserved on average over a Hubble time.
This is a similar situation as the one with hydrogen recombination: direct recombinations to the $1s$ state of atomic hydrogen do not contribute to the recombination process, because the 13.6 eV photons emitted will soon ionize newly-formed atoms. In the case of hydrogen,  $2s-1s$ two-photon processes are instead needed in order for cosmic recombination to proceed. 

We do not consider here the existence of excited states of the DM particle, but will instead note that the photon emitted after the capture possibly quickly interacts and thermalizes with the cosmological plasma, and is practically ``lost'' for the purpose of the inverse process. If such a regime is realized, the only photons that are available for the inverse process are those of the cosmic background. 

We thus first proceed to check when this regime is realized, and then, armed with that information, study the evolution of the DM and baryon abundances using standard tools (e.g. the Saha equation). We will focus on redshifts before hydrogen recombination, i.e., $z>1100$.
We compare the interaction rate of the emitted photon with the primordial plasma $\Gamma_{\gamma \text{Pl}}$ to the Hubble expansion rate $H$, which is given by
\begin{equation}
H(z) = H_0 \left[\Omega_r (1+z)^4 + \Omega_m (1+z)^3 + \Omega_{\Lambda}\right]^{1/2} \, . \label{eq:hubblerate}
\end{equation} 
The rate $\Gamma_{\gamma \text{Pl}}$ can be written as the sum of different contributions
\begin{equation}
\Gamma_{\gamma \text{Pl}} = \Gamma_{\text{Comp}} + \Gamma_{\text{PS}} + \Gamma_{\text{PPn}} + \Gamma_{\text{PP}\gamma} \, , \label{eq:intgplasma}
\end{equation}
consisting of Compton scattering, photon scattering, pair production over nuclei (both H and $^{4}$He) and pair production over photons, respectively. The explicit expression for each of these contributions can be found in Appendix \ref{App:int-rates}.
The other relevant quantity is the interaction rate of emitted photons with macros, $\Gamma_{\gamma X}$:
\begin{equation}
\Gamma_{\gamma X} = n_X \sigma_X = \Omega_\text{DM} \, \rho_{c,0} \,\frac{\sigma_X}{M_X}\, (1+z)^3 \, ,
\end{equation}
where the cross-section $\sigma_X$ is purely geometrical\footnote{Since we are considering macroscopic dark matter candidates, which have a radius $R_X$ much larger than any relevant microscopic length scale, we are ignoring the quantum-mechanical aspects of the interaction of Macros with other particles (see \cite{Jacobs:2014yca}). As a result, the interaction cross-sections of Macros with both photons and baryons are purely geometrical.}, i.e. $\sigma_X = \pi R_X^2$, $M_X$ is the mass of Macros $\Omega_\text{DM}$ is the present DM density and $\rho_{c,0}$ is the current value of the critical density of the Universe. Notice that we are also assuming that the dark matter is entirely made up of Macros, such that $n_X \equiv n_{\text{DM}}$.

\begin{figure}
	\centering
	\includegraphics[scale=0.348]{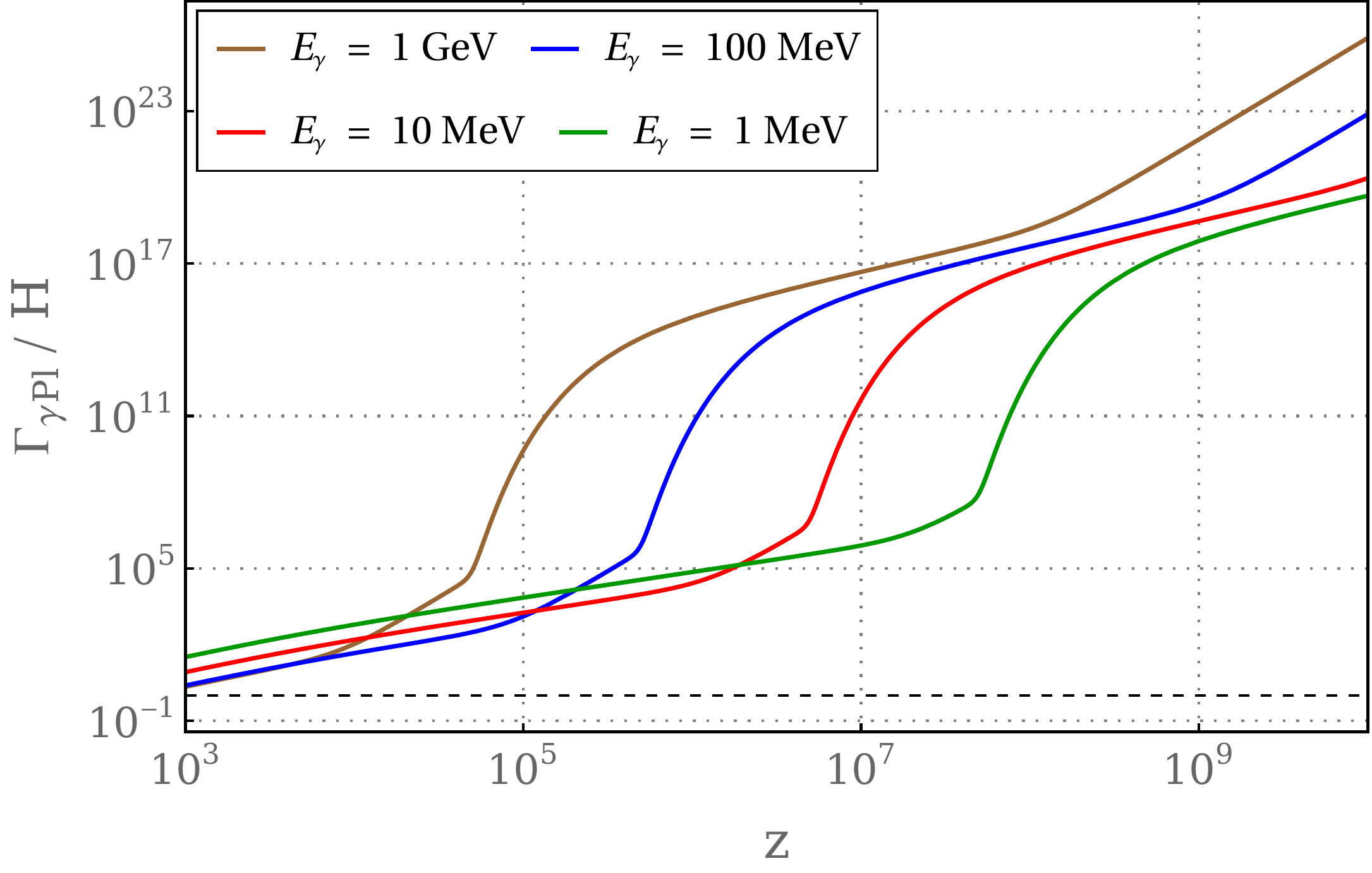}
	\includegraphics[scale=0.348]{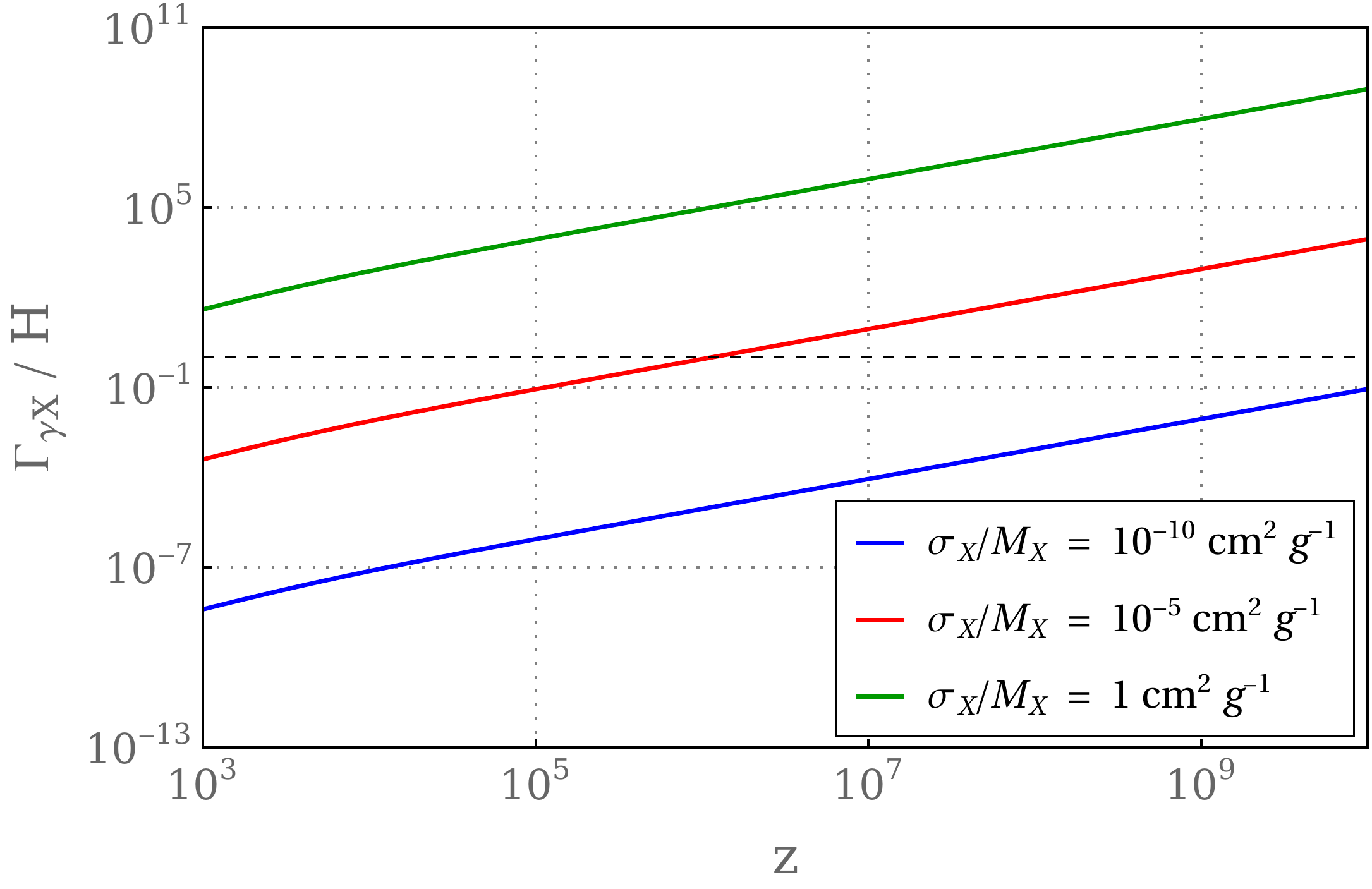}
	\includegraphics[scale=0.38]{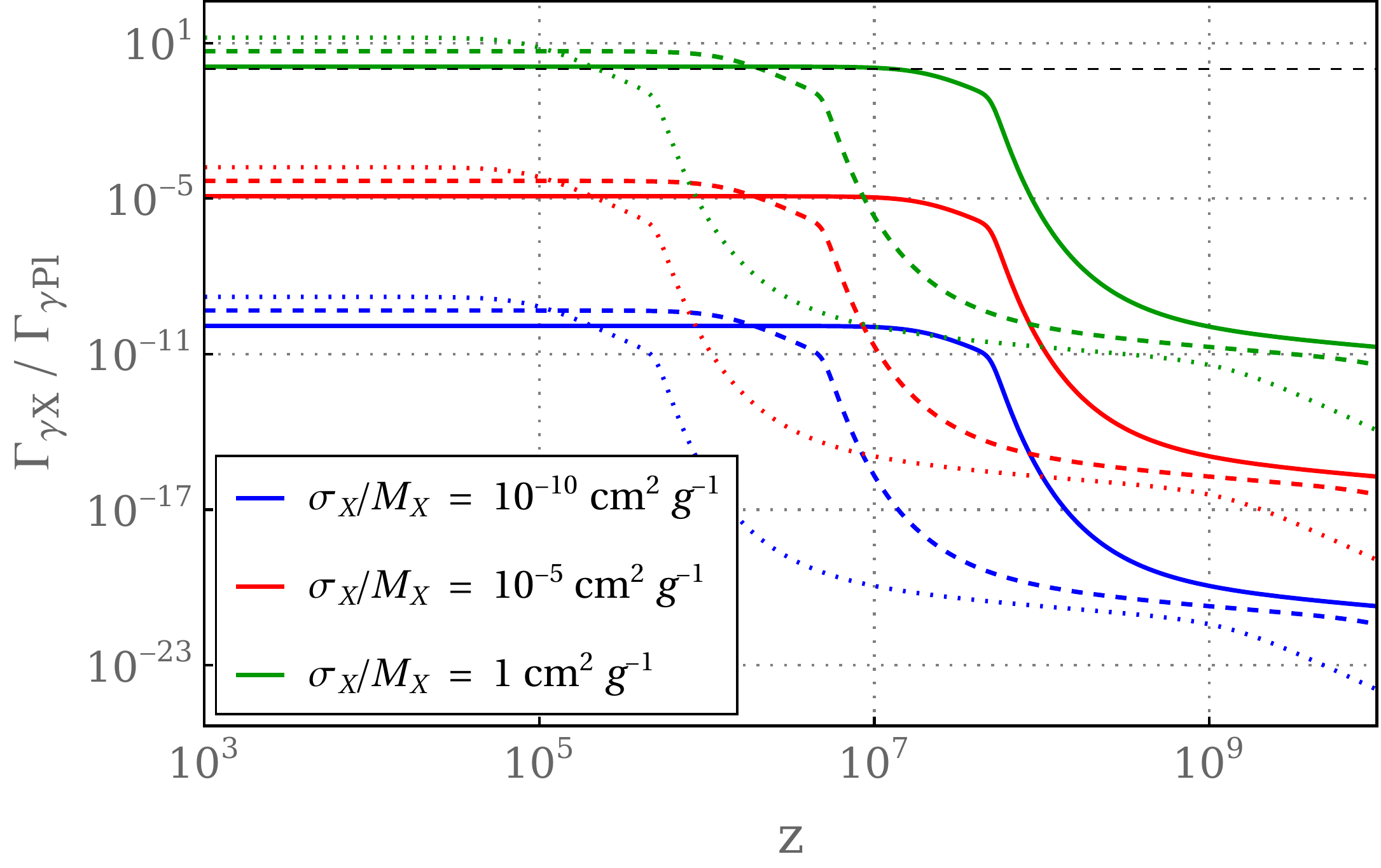}
	\caption{On the left we report the interaction rate of photons with the primordial plasma over the Hubble expansion rate, for different energy values: $1$ MeV (green), $10$ MeV (red), $100$ MeV (blue) and $1$ GeV (brown). On the right  we show the interaction rate of the emitted photon with Macros over the Hubble rate, for different values of the reduced cross-section.
	On the bottom panel, we report the ratio between the interaction rate of the emitted photon with Macros and with the primordial plasma, for different values of the reduced cross-section and of the energy of photons: $1$ MeV (solid), $10$ MeV (dashed) and $100$ MeV (dotted). The black dashed lines represent the condition $\Gamma_{\gamma\text{Pl}}/H = 1$, $\Gamma_{\gamma X}/H = 1$ and $\Gamma_{\gamma X}/\Gamma_{\gamma\text{Pl}} = 1$, respectively.}
	\label{fig:intpl-Hub}
\end{figure}

In Figure \ref{fig:intpl-Hub} we compare these three rates, for different values of the injected photon energy $E_{\gamma}$ and of the reduced cross section $\sigma_X/M_X$, as a function of redshift $z$. In the upper right panel, the ratio between $\GgX$ and the Hubble parameter is shown. For values of the reduced cross-section $\sM > 10^{-1} \, \cmsq\,\gram^{-1}$, interactions between photons and macros are frequent until after recombination.
For values $\sM < 10^{-8} \, \cmsq\,\gram^{-1}$, interactions become irrelevant. 
The upper left panel shows that $\Ggpl > H$ for the energies and redshifts under consideration. Note that this ratio does not depend on $\sigma_X/M_X$. The bottom panel finally shows that $\Gamma_{\gamma X} < \Ggpl$ for $\sigma_X / M_X \lesssim 1 \, {\text{cm}^2 \, \text{g}^{-1}}$. These facts together imply that, in this region of parameter space, the injected photons thermalize efficiently with the primordial plasma before having the chance to interact with a Macro. In this regime, the chemical equilibrium of the reaction \eqref{eq:bcapture} is regulated uniquely by the cosmic background photons. In the opposite regime, $\sigma_X / M_X \gtrsim 1 \, {\text{cm}^2 \, \text{g}^{-1}}$, the absorption of the injected photons by Macros becomes more efficient than the absorption by the primordial plasma, since $\Gamma_{\gamma X} > \Ggpl$. When this is realized, the comoving density of baryons is again conserved. In the following, we will focus on the former case\footnote{Strictly speaking, the comoving density is constant for $\Gamma_{\gamma X} \gg \Ggpl$. 
When the two rates are comparable, $\Gamma_{\gamma X} \simeq \Ggpl$, a fraction of protons will still be absorbed by Macros, possibly leaving some signatures in the cosmological observables. However, we make the conservative assumption that our analysis is valid only for $\sigma_X / M_X < 1 \, {\text{cm}^2 \, \text{g}^{-1}}$.}, namely $\sigma_X / M_X \lesssim 1 \, {\text{cm}^2 \, \text{g}^{-1}}$.

Macros are kept in equilibrium with the cosmological plasma as long as the rate for proton capture is faster than the expansion rate.
The rate  at which macros $X^N$ absorb baryons is given by:
\begin{equation}
\Gamma_\mathrm{abs} = n_b \sigma_X v_\mathrm{rel} \simeq 21.2 \left( \frac{\Omega_b h^2}{0.022}\right)\left(\frac{\sigma_X}{10^{-15}\, \cm^2}\right)
\left(\frac{T}{\MeV}\right)^{7/2} \,\eV \,,
\end{equation}
where $v_\mathrm{rel}$ is the relative velocity between the macro and baryon fluids, $\Omega_b h^2$ is the physical baryon density and $T$ denotes the temperature of the baryon-photon plasma. In the last equality, we have used the fact that, when the fluids are coupled, the relative velocity is given
by the thermal velocity of baryons, $v_\mathrm{rel}\simeq v_\mathrm{th} = \sqrt{3T/m_p}$.

Deep into the radiation- and matter-dominated eras the Hubble rate can be written as:
\begin{equation}
H = \left\{ \begin{array}{ll}
2.50\times 10^{-16} \left(\frac{T}{\MeV}\right)^2\,\eV\, , & \qquad \mathrm{(RD)} \\[0.2cm]
2.05\times 10^{-19} \left(\frac{\Omega_\text{DM} h^2}{0.12}\right)^{1/2}\left(\frac{T}{\MeV}\right)^{3/2}\,\eV & \qquad \mathrm{(MD)}
\end{array}\right.
\end{equation} 
From the above expressions, it is pretty straightforward to check that Macros stay in equilibrium until after recombination ($z=1100$) if the capture cross section $\sigma_X \gtrsim 10^{-22} \,\cm^2$.

At equilibrium, the densities of $X^N$'s and $X^{N+1}$'s are related by the Saha equation (valid for $T\ll m_B$): 
\begin{equation}
f_X \equiv \frac{n_X^{N+1}}{n_X^{N}} = \frac{2^{5/2}\zeta(3)}{\sqrt{\pi}} \, \epsilon_B \left(\frac{g_{N+1}}{g_N \, g_B}\right) \eta \left(\frac{T}{m_B}\right)^{3/2} e^{I/T} \, , \label{eq:saha}
\end{equation}
where $\epsilon_B = n_B / n_b$ is the fraction of baryons in protons or neutrons\footnote{After the Big Bang Nucleosynthesis (BBN), i.e. at temperatures $T \lesssim 0.1$ MeV, there are no free neutrons left in the Universe and we have 
	\begin{equation}
	\epsilon_p = \frac{1+2f_{\text{He}}}{1+4f_{\text{He}}} \, , \quad \epsilon_n = 0 \, ,
	\end{equation}
	where $f_{\text{He}} \simeq Y_p/[4 (1 - Y_p) ]$ and $Y_p$ is the primordial Helium mass fraction. At temperatures $T \gtrsim 1$ MeV, instead, weak interactions maintain the balance between neutrons and protons and thus
	\begin{equation}
	\epsilon_p = \epsilon_n = \frac{1}{2} \, .
	\end{equation}
}, $I$ is the binding energy of macros, $\eta = n_b / n_\gamma = 5.5 \cdot 10^{-10} (\Omega_bh^2/0.02)$ is the baryon-to-photon ratio, $m_B$ is the mass of the neutron or proton and $g$ is the number of internal degrees of freedom. As regards the latter, $g = 2$ for neutrons and protons, while we assume that $g_{N+1} / g_N \simeq 1$ for MDM. When deriving the Saha equation, we have assumed that both protons and photons are in thermal equilibrium. The equilibrium for protons is justified by the fact that, as we shall see in the next section, observations allow for only a small fraction of protons to be absorbed by Macros. In the case of photons, we are instead using the fact that the high-energy photons released after the capture quickly thermalize with the plasma, as commented above.

\begin{figure}
	\centering
	\includegraphics[scale=0.5]{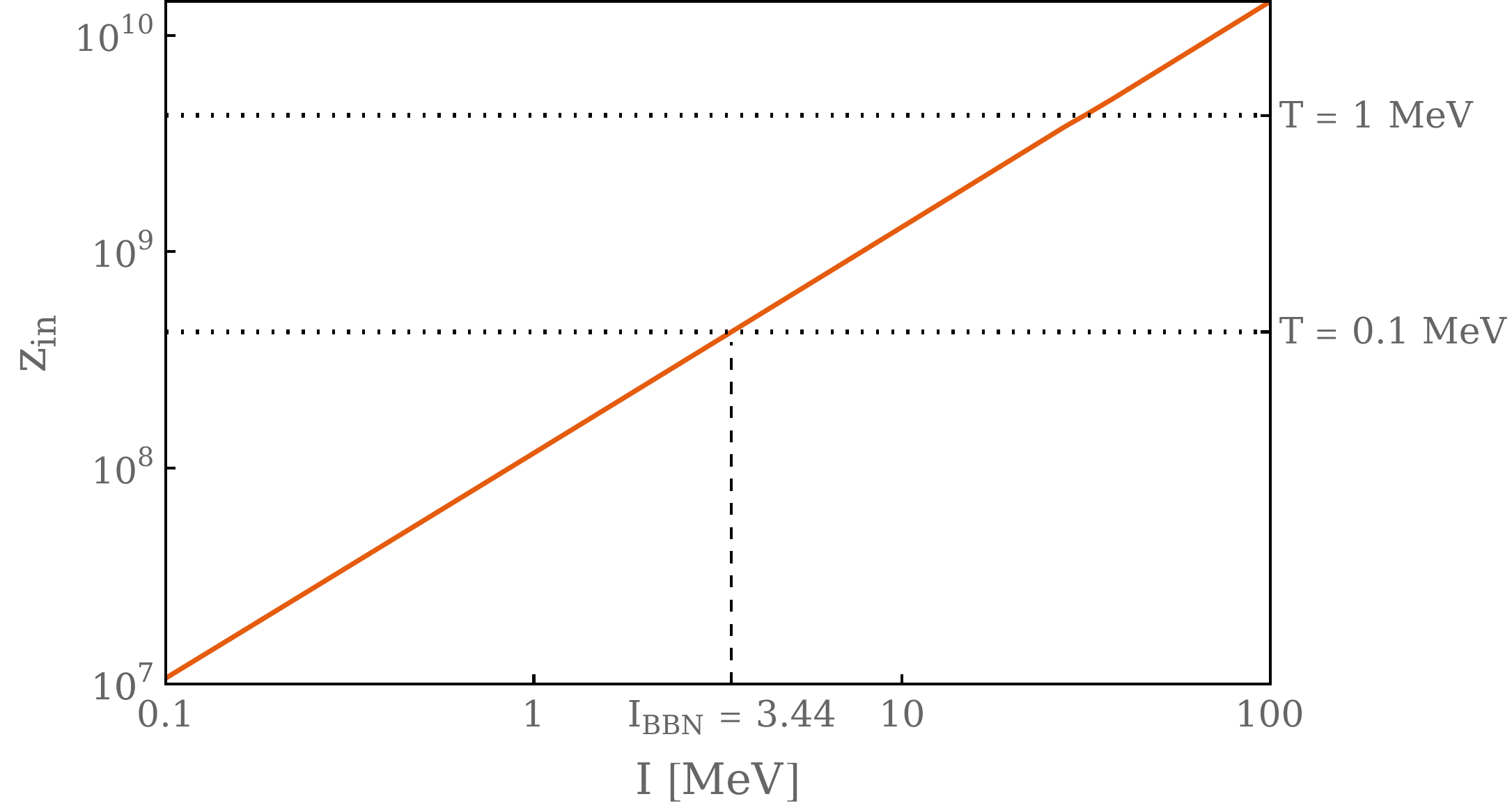}
	\caption{Redshift at which the reaction \eqref{eq:bcapture} goes out of equilibrium (in the case of protons) as a function of the Macro binding energy $I$. We define a threshold value of the binding energy, $I_{\text{BBN}} \simeq 3.44$ MeV, such that for $I < I_{\text{BBN}}$ the absorption of baryons starts after the BBN.}
	\label{fig:zin}
\end{figure}

From the Saha equation, it is seen that $f_X \ll 1$ for $T\gg I$, i.e., only the $X^N$'s are populated. Hence there is no net absorption of baryons. However, as the Universe expands and cools down, the abundance of $X^N$'s starts to be Boltzmann suppressed since there are fewer photons energetic enough to photodissociate the $X^{N+1}$. When this happens, macros effectively start absorbing baryons. As a benchmark, we will take the redshift $z_{\text{in}}$ at which proton capture starts as the one when $f_X = 1$. 
In Figure \ref{fig:zin} we plot $\zin$ as a function of the Macro binding energy $I$.
As expected, the higher is the binding energy $I$, the higher the redshift at which photodissociation of Macros becomes ineffective. In particular, there is a threshold value of the binding energy, $I_{\text{BBN}} \simeq 3 \,\MeV$, such that for $I \ge I_{\text{BBN}}$ the effective absorption of baryons starts before Big Bang Nucleosynthesis (BBN) is complete (we take BBN to end at $T \simeq 0.1$ MeV). In this case, both protons and neutrons are absorbed by Macros and this might affect the standard BBN picture, since the interaction rates of the two particle species with Macros are in general different (as discussed in Ref.~\cite{Jacobs:2014yca}). Moreover, the high-energy photons that are emitted during the process might also affect the production of light elements.
In the opposite regime, i.e. when $I < I_{\text{BBN}}$, the process \eqref{eq:bcapture} starts after the end of the BBN. 
Since there are no free neutrons left in the Universe after BBN, Macros interact only with protons. In the following, we will concentrate on the latter case.

\subsection{Baryon density between BBN and CMB epochs}
\label{subsec:baryon}

Cosmological observations allow to measure the baryon density at different epochs. Light element abundances depend on the baryon density at the time of BBN, while CMB anisotropies probe this quantity around the time of hydrogen recombination\footnote{More correctly, light element abundances allow to probe the baryon-to-photon ratio at the time of BBN. Things are more complicated for the CMB, since the anisotropy pattern depends on both $\eta$ and $n_b$ around the time of recombination. However, since we are neglecting the effect of photons injected between BBN and recombination, we can drop the distinction.}. 
For the BBN we take the value of $\Omega_b h^2$ derived in \cite{Consiglio:2017pot}, which has been obtained using the measurements of the abundances of deuterium and $^4$He from Refs. \cite{Cooke:2017cwo} and \cite{Aver:2015iza}, respectively\footnote{Recently, a new analysis has been performed in \cite{Gariazzo:2021iiu} using updated expressions for nuclear rates. The inclusion of these new results does not alter significantly our conclusions.}. For the CMB, instead, we use the constraint from the 2018 Planck data release \cite{Aghanim:2018eyx}:
\begin{equation}
(\Omega_b h^2)_{\text{BBN}} = 0.0227 \pm 0.0005 \, , \quad (\Omega_b h^2)_{\text{CMB}} = 0.02236 \pm 0.00015 \, .
\label{eq:omegab_obs}
\end{equation} 
These values are consistent within the respective uncertainties. This fact can be used to constrain the amount of protons captured by Macros between BBN and recombination, and from that the capture cross-section.

Let us define the comoving number density of protons as\footnote{In our analysis we are assuming that only free protons of the cosmological plasma are captured by Macros, neglecting instead the possibility that He nuclei are also absorbed.} 
$\mathcal{N}_p \equiv a(t)^3 \, n_p $.
Following the discussion in the previous section, we can take $\mathcal{N}_p = \mathrm{const}$ for $z>\zin$. At lower redshifts, we can neglect the photodissociation of Macros and the comoving density evolves according to
\begin{equation}
\dot{\mathcal{N}}_p = -\Gamma_{pX} \, \mathcal{N}_p \qquad(z<\zin) \, , \label{eq:proton-ev} 
\end{equation}
where $\Gamma_{pX}$ is the rate at which protons are captured by Macros.
For Macros with surface potential $V_X$, the capture rate can be written as \cite{Jacobs:2014yca}
\begin{equation}
\Gamma_{pX} = n_X \sigma_X v_{\text{rel}} \cdot
\begin{cases}
e^{- V_X / T} \qquad\qquad\, V_X \ge 0 \, , \\
1 - V_X / T \qquad\quad V_X < 0 \, ,
\end{cases} \label{eq:rate-p}
\end{equation}
where we approximate the relative velocity between Macros and protons $v_{\text{rel}}$ with the thermal velocity $v_\mathrm{th}$ of the latter (see discussion in Sec.~\ref{sec:Tight-coupling}).
We will mainly consider values of the surface potential $|V_X| \simeq \mathcal{O}(\mathrm{MeV})$, as expected for strangelets and nuclear-type MDM \cite{Alcock:1986hz}. However, to keep our discussion more general, we will also analyze cases with smaller values of $V_X$.
If we think of nuclear-like Macros, we expect them to have a positive electric charge. In this case, as discussed e.g. in \cite{Alcock:1986hz} for strange matter, an external shell of electrons is formed around the Macro such that the global charge vanishes, while protons of the cosmological plasma feel an effective negative charge when they are close to the external electronic shell.

The temperature of the baryon-photon plasma evolves as $T \simeq T_0 (1+z)$, where $T_0$ denotes the present-day CMB temperature.
As above, we use the fact that
\begin{equation}
n_X \sigma_X = \Omega_\text{DM} \, \rho_{c,0} \,\frac{\sigma_X}{M_X}\, (1+z)^3 \, ,
\end{equation}
when evaluating the interaction rate.

If Macros have a positive surface potential, protons have to face a potential barrier which tends to suppress their absorption rate, as encoded in the exponential factor present in Eq. \eqref{eq:rate-p}. For negative surface potentials, the proton absorption rate is instead enhanced.

The formal solution to Eq. \eqref{eq:proton-ev} reads, for $t>t_\mathrm{in}$,
\begin{equation}
\mathcal{N}_p (t) = \mathcal{N}_{p,\text{in}} \, \exp \left[ - \int_{t_{\text{in}}}^t \Gamma_{pX} \, dt \right] \, , \label{eq:proton-sol} \\
\end{equation}
where $t_\mathrm{in}$ represents the time at which the absorption of protons by Macros starts, and  $\mathcal{N}_{p,\text{in}} = \mathcal{N}_p (t_\mathrm{in})$. The logarithmic ratio of the comoving number densities at BBN and recombination is then:
\begin{equation}
\Delta \log{\mathcal N}_p \equiv \log\left(\frac{\mathcal{N}_p (t_{\mathrm{BBN}})}{{\mathcal{N}_p (t_\mathrm{CMB})}}\right) = \int_{\max(t_\mathrm{in},t_\mathrm{BBN})}^{t_\mathrm{CMB}} \Gamma_{pX} \, dt \, . \label{eq:bcapture-constraints} 
\end{equation}
This expression can be used to evaluate $\Delta \log{\mathcal N}_p$ for given values of $\sigma_X/M_X$, $V_X$ and $I$ (since the latter determines $\zin$).

\begin{figure}
	\centering
	\includegraphics[scale=0.65]{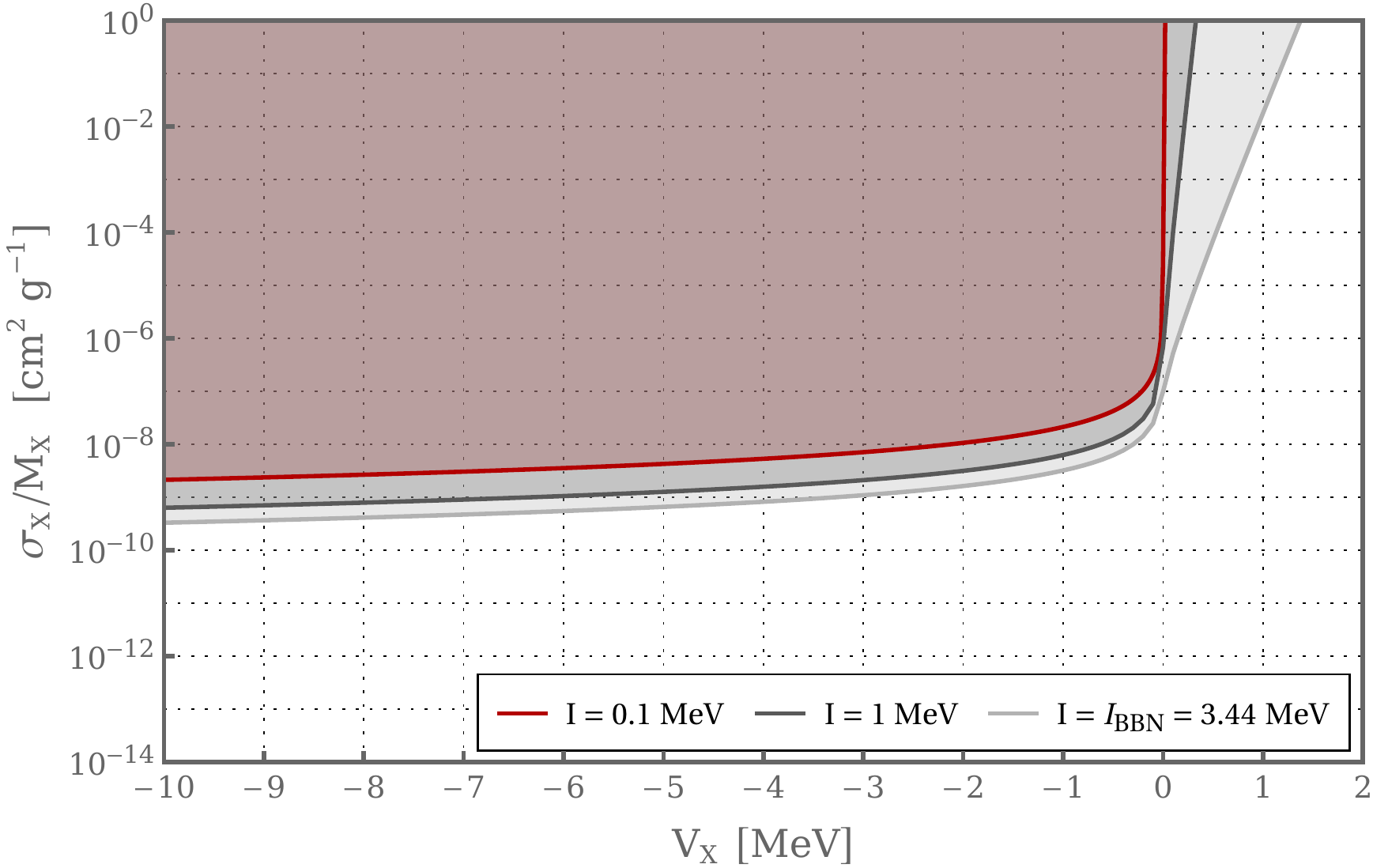}
	\caption{Constraints on the reduced cross-section as a function of the Macro surface potential, for different values of the binding energy $I$. The shaded regions are excluded by the BBN and CMB measurements of the baryon density. As discussed in Section \ref{subsec:Thermodynamics}, these constraints are valid for $\sigma_X/M_X \lesssim 1 \, \mathrm{cm}^2 \, \mathrm{g}^{-1}$, since for higher values of the reduced cross-section the injected photons are not absorbed efficiently by the cosmological plasma.}
	\label{fig:const_baryon_numb}
\end{figure}

Using the measured values in Eq. (\ref{eq:omegab_obs}), we find the following observational constraint:
\begin{equation}
(\Delta \log{\mathcal N}_p)_\mathrm{obs} = 0.015 \pm 0.029 \,. \label{eq:Delta_Np}
\end{equation}
Enforcing this constraint produces the exclusion plot shown in Figure~\ref{fig:const_baryon_numb}.

\subsection{CMB spectral distortions}

The capture of protons by Macros leads also to the injection of high-energy photons in the primordial plasma, see Eq. \eqref{eq:bcapture}. This might produce spectral distortions in the CMB \cite{Hu:1992dc,Chluba:2011hw,Chluba:2013vsa,Chluba:2016bvg,Kogut:2019vqh,Chluba:2019nxa}.

If energy is released in the baryon-photon plasma during the so-called $\mu$-era, i.e. between the thermalization redshift
\begin{equation}
z_ {\text{th}} \approx 1.98 \cdot 10^6 \left(\frac{1 - Y_p / 2}{0.8767}\right)^{-2 / 5}\left(\frac{\Omega_b h^2}{0.02225}\right)^{-2 / 5} \left(\frac{T_0}{2.726 \, \text{K}}\right)^{1/5}
\end{equation} 
and $z = z_{\mu y} \simeq 5 \cdot 10^4$, Compton scattering will still be able to drive the plasma towards kinetic equilibrium. However, the photon number-changing processes, like Bremsstrahlung and double Compton, are inefficient in this range of redshifts (see e.g. \cite{danese1982double,Hu:1992dc}). Because of this, energy injection in the $\mu$-era makes CMB photons acquire a Bose-Einstein distribution function with chemical potential $\mu \ne 0$. This deviation from a perfect blackbody spectrum is referred to as $\mu$-distortion. At redshift $z \simeq z_{\mu y}$, instead, thermalization by Compton scattering becomes also inefficient and the transition between $\mu$ and $y$-distortions takes place. 

\begin{figure}
	\centering
	\includegraphics[scale=0.45]{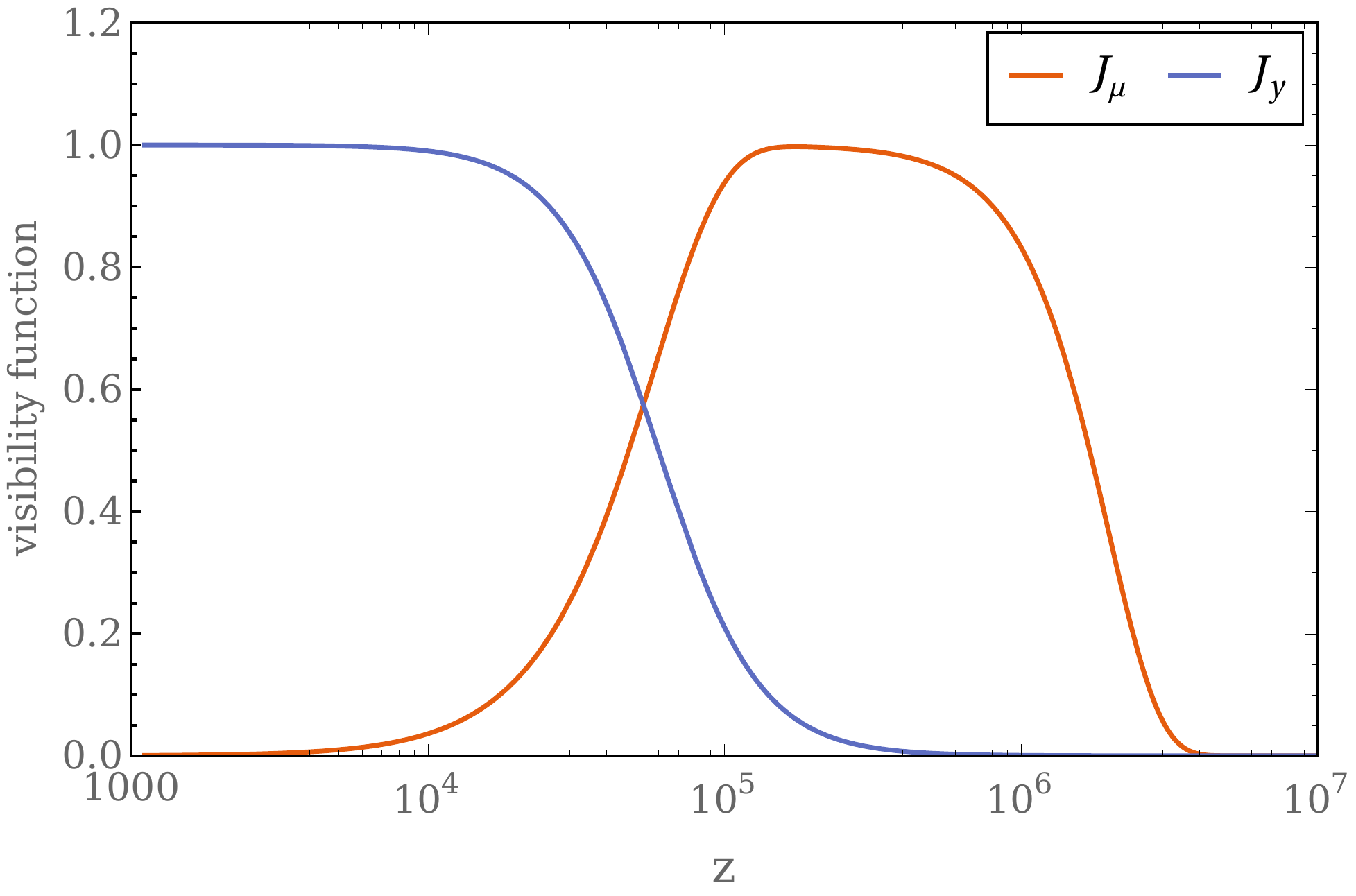}
	\caption{Visibility functions for $\mu$ and $y$-distortions as a function of the redshift $z$. 
	It can be seen that energy injection taking place at redshifts $10^4 \lesssim z \lesssim 3 \cdot 10^5$ results in a superposition of $\mu$ and $y$-types.}
	\label{fig:visibility}
\end{figure}

The $\mu$ and $y$-distortions can be computed as (see e.g. \cite{Chluba:2011hw,Chluba:2013vsa,Chluba:2016bvg})
\begin{align}
\mu &\simeq 1.401 \int^{z_{\text{in}}}_0 \mathcal{J}_{\mu}(z) \frac{\dot{Q}}{\rho_\gamma (z)} \frac{dz}{H(z) (1+z)} \, , \label{eq:mu-distorsions} \\[5pt] 
y &\simeq \frac{1}{4} \int^{z_{\text{in}}}_0 \mathcal{J}_y(z) \frac{\dot{Q}}{\rho_\gamma (z)} \frac{dz}{H(z) (1+z)} \, , \label{eq:y-distorsions}
\end{align}
where $z_{\text{in}}$ is the redshift at which the energy release begins (in our case, when photodissociation of Macros becomes ineffective), $\rho_{\gamma} = (\pi^2/15) T^4$ is the energy density of background photons, $\dot{Q}$ is the heating rate and $\mathcal{J}_{\mu/y}(z)$ are the distortion visibility functions, which quantify the fraction of the energy injected into the baryon-photon plasma that contributes to $\mu$ and $y$-distortions, respectively. These can be analitically approximated as \cite{Chluba:2013vsa,Chluba:2016bvg}
\begin{align}
\mathcal{J}_{\mu}(z) &\approx e^{-(z/z_{\text{th}})^{5/2}} \Bigg\{ 1 - \exp \left[- \left(\frac{1+z}{5.8 \cdot 10^4}\right)^{1.88} \right] \Bigg\} \, , \label{eq:mu-visibility} \\[8pt]
\mathcal{J}_y(z) &\approx
\begin{cases}
\left[ 1 + \left(\frac{1+z}{6 \cdot 10^4}\right)^{2.58} \right]^{-1} \quad\; z \ge z_\text{rec} \vspace{2mm}\\
0 \qquad\qquad\qquad\qquad\quad z < z_\text{rec} \, .
\end{cases} \label{eq:y-visibility}
\end{align}
The contribution $e^{-(z/z_{\text{th}})^{5/2}}$ in $\mathcal{J}_{\mu}(z)$ accounts for the fact that even at redshifts $z > z_{\text{th}}$ a small amount of $\mu$-distortions is produced. The other two terms in $\mathcal{J}_{\mu}(z)$ and $\mathcal{J}_y(z)$, instead, account for the fact that the transition between $\mu$ and $y$-distortions is not sudden at $z = z_{\mu y}$. Rather, around this redshift the distortions consist of a superposition of $\mu$ and $y$-types. This can be seen in Figure \ref{fig:visibility}, where the distortion visibility functions for both $\mu$ and $y$-types are shown as a function of the redshift. 

The heating rate $\dot{Q}$ is in our case given by
\begin{equation}
\dot{Q} = n_p(z) \Gamma_{p X}(z) I \, ,
\end{equation}
where the time evolution of the proton density is computed using Eq. \eqref{eq:proton-ev} and hence taking into account the capture by Macros.
Notice that we are assuming that all the energy released goes into heating. This is justified because, as we have shown in Section \ref{subsec:Thermodynamics}, the interaction rate of the emitted photons with the primordial plasma is always much greater than the Hubble expansion rate. 

\begin{figure}
	\centering
	\includegraphics[scale=0.66]{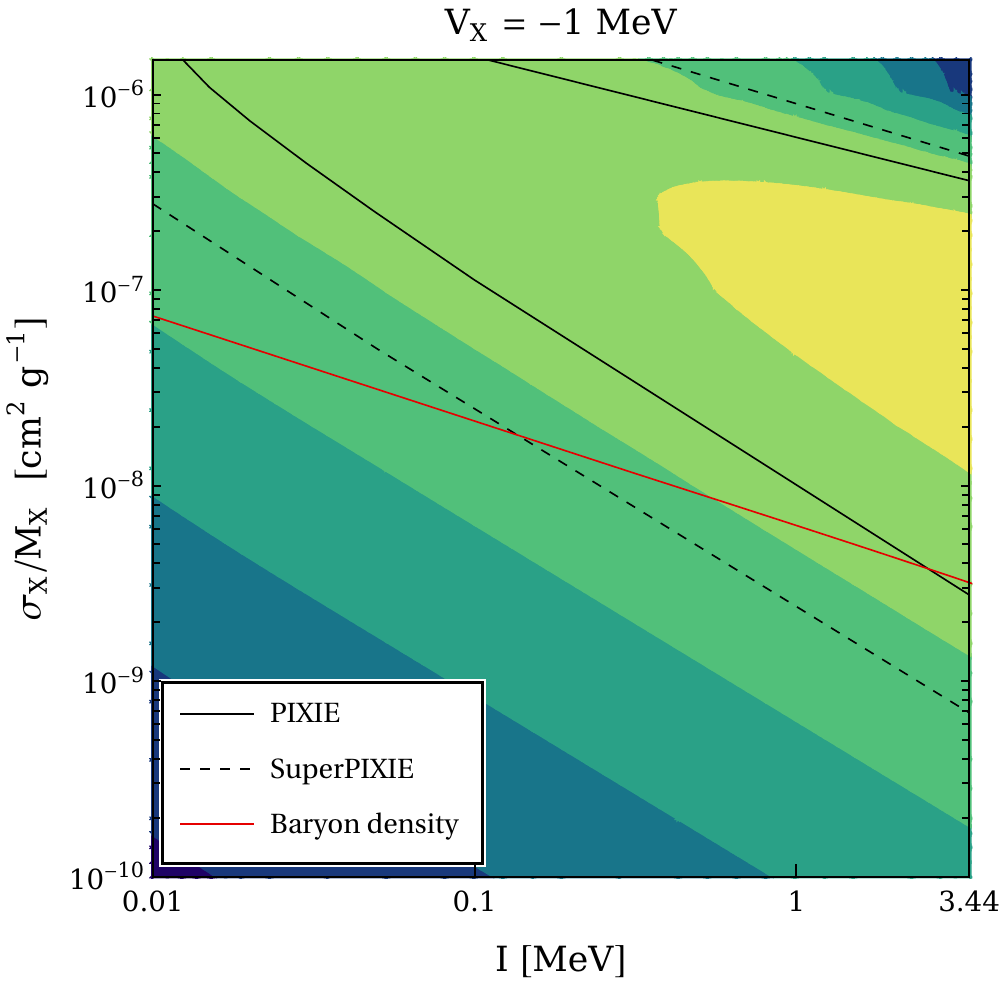}
	\includegraphics[scale=0.66]{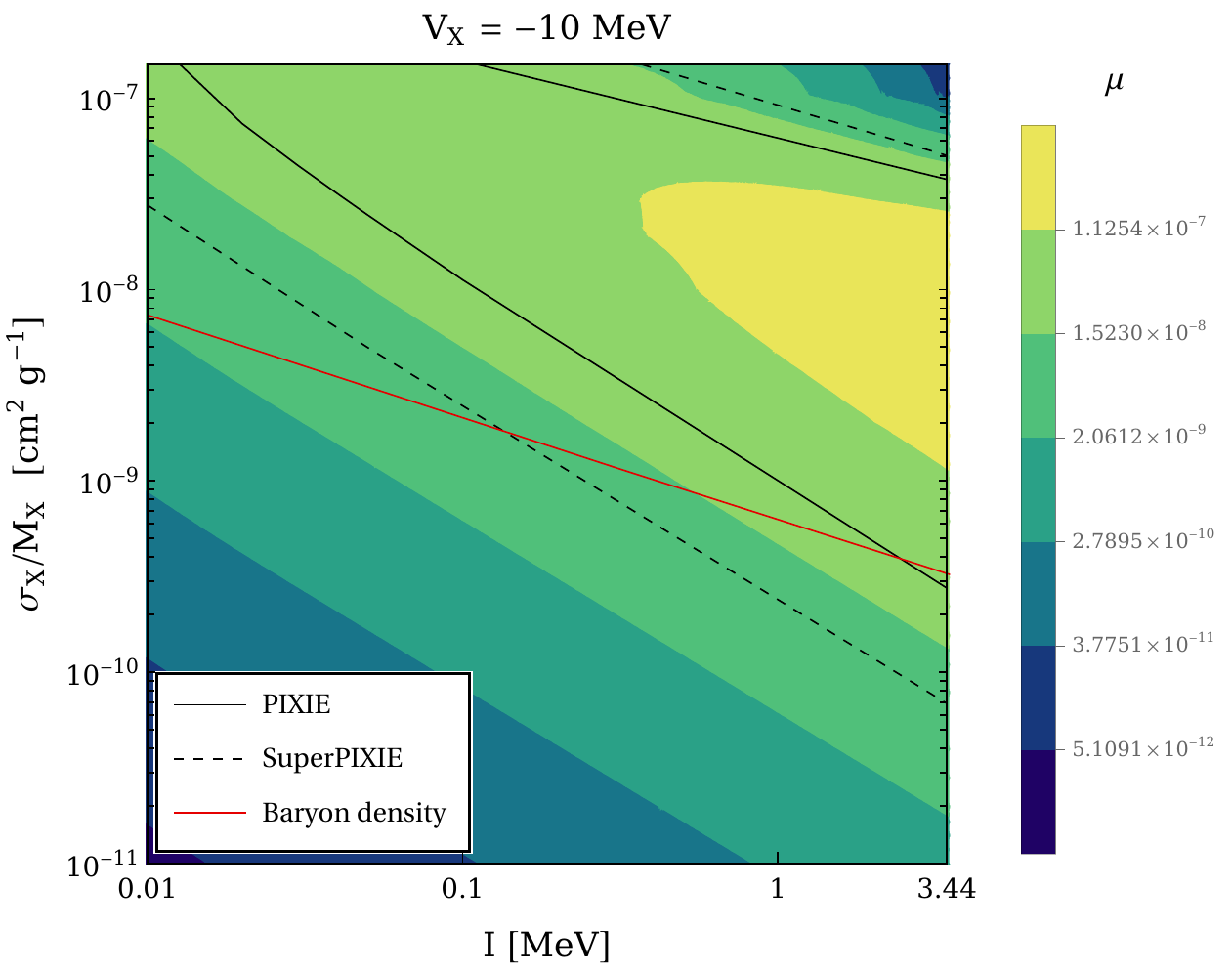}\vspace{2mm}
	\includegraphics[scale=0.66]{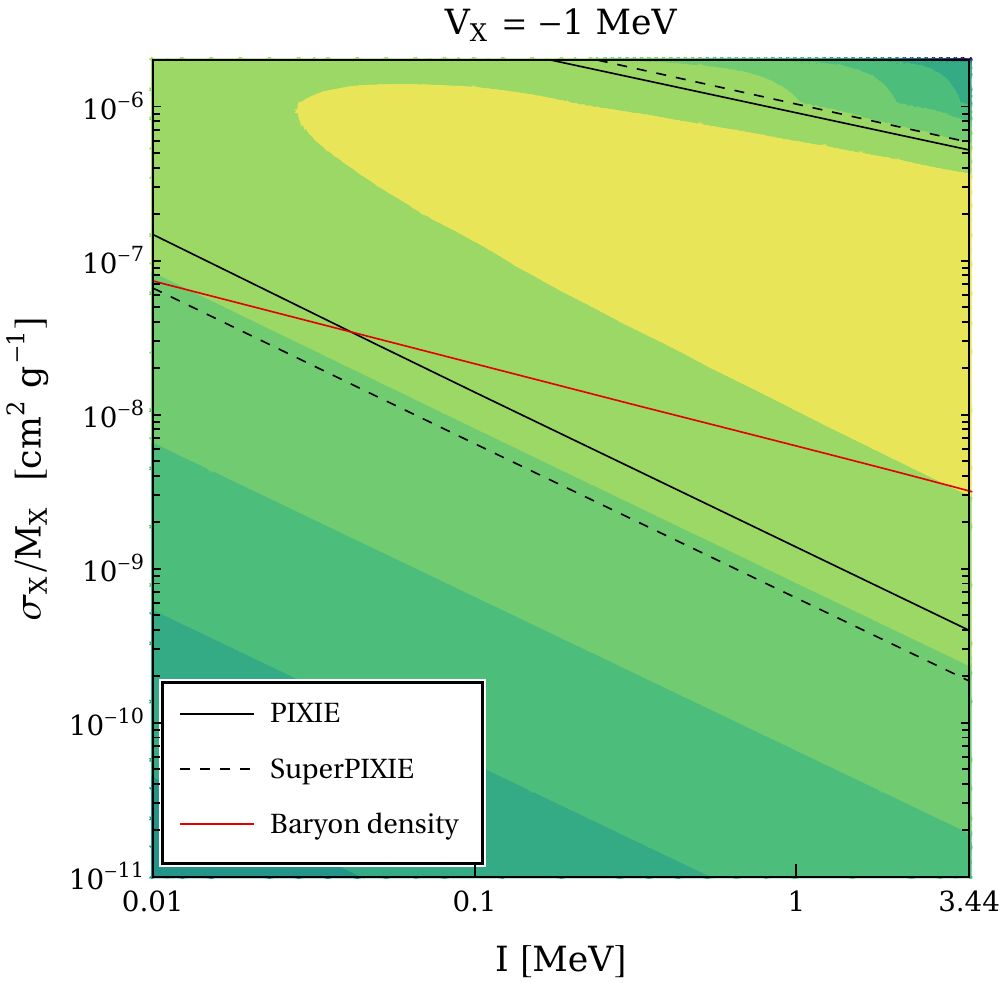}
	\includegraphics[scale=0.66]{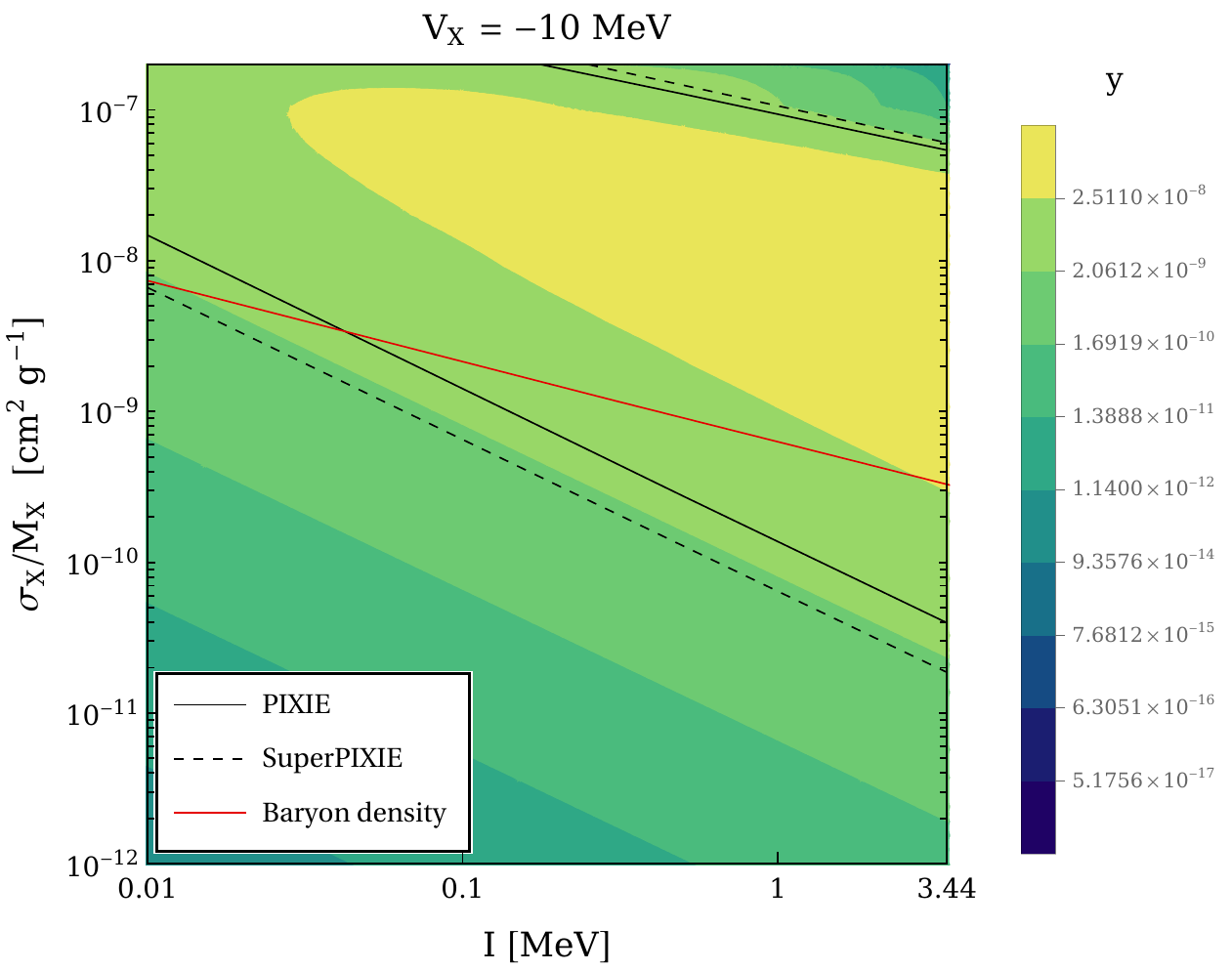}
	\caption{Contour plots of $\mu$ and $y$-distortions as a function of the reduced cross-section $\sigma_X / M_X$ and the Macro binding energy $I$, for two different values of the surface potential $V_X$: $-1$ MeV (left panel) and $-10$ MeV (right panel). The regions of parameter space within the black solid and dashed lines are excluded assuming a null detection of spectral distortions from PIXIE ($|\mu| < 3 \cdot 10^{-8}$, $|y| < 3.4 \cdot 10^{-9}$) and SuperPIXIE ($|\mu| < 7.7 \cdot 10^{-9}$, $|y| < 1.6 \cdot 10^{-9}$), respectively. The regions above the red lines are excluded by the constraints from the baryon density, as discussed in the previous section.}
	\label{fig:distortions-contour}
\end{figure}

In Figure \ref{fig:distortions-contour} we show the amount of $\mu$ and $y$-distortions that are produced as a function of the parameters of the model. Both types of spectral distortions are always below the current upper bounds set by FIRAS, $|\mu| < 9 \cdot 10^{-5}$ and $|y| < 1.5 \cdot 10^{-5}$ (95\% C.L.) \cite{Mather:1993ij,Fixsen:1996nj}. However, they are within the reach of the proposed PIXIE and SuperPIXIE experiments which, in case of no detection, would set the $1 \sigma$ limits $|\mu| < 3 \cdot 10^{-8}$, $|y| < 3.4 \cdot 10^{-9}$ \cite{Kogut:2011xw,Chluba:2019nxa} and $|\mu| < 7.7 \cdot 10^{-9}$, $|y| < 1.6 \cdot 10^{-9}$ \cite{Kogut:2019vqh,Chluba:2019nxa}, respectively. The region of the parameter space that would be accessible to PIXIE (SuperPIXIE) are those within the solid (dashed) lines. 

Notice that the amount of spectral distortions is maximized for values of the reduced cross-section $\sigma_X/M_X \sim 10^{-6}$ and $10^{-7}$ for $V_X = -1$ MeV and $V_X = -10$ MeV, respectively. As one might naively expect, spectral distortions decrease for smaller values of the reduced cross-section since fewer protons are absorbed, resulting in less photons being released in the baryon-photon plasma. However, the amount of spectral distortions decreases also for higher values of the reduced cross-section: this happens because a large number of protons is absorbed at redshifts close to $z_\text{in}$; thus, for high values of $I$ (and hence of $z_\text{in}$) most of the energy release takes place before the $\mu$ and $y$-eras, thus not giving rise to any detectable spectral distortion.

Finally, let us remark that no sizeable spectral distortions are produced if $V_X \ge 0$. Indeed, as we can see from Figure \ref{fig:const_baryon_numb}, only in the case with $I \sim I_{\text{BBN}}$ a non-negligible amount of protons is absorbed if these have to face a potential barrier. However, also in this case most of the absorptions take place at high redshifts, when the thermal energy of protons is large enough to overcome the potential barrier.

\subsection{Tight coupling between baryons and charged Macro DM at recombination}
\label{sec:Tight-coupling}

If the dark matter is electrically charged, it can scatter off electrons and protons with the possibility of getting coupled to the baryon-photon plasma. If this condition were realized at recombination, MDM would effectively behave like baryons as regards its effects on the CMB anisotropies. Since we are assuming that Macros form the entirety of the DM, this can not be the case and hence we must require that the two fluids are not coupled at recombination\footnote{In this discussion we are assuming that the dark matter fluid approximation is still valid, in spite of  the low number density of Macros. This is true provided that the diffusion time for a photon to cross the average separation between two Macros is small compared to the Hubble time, $H^{-1}$. In terms of the mass of Macros, this condition can be written as \cite{Sidhu:2019gwo}
\begin{equation}
\left( \frac{M_X}{\text{g}} \right)^{2/3} (1+z)^3 \ll 10^{39} \, .
\end{equation}  
Even for very massive MDM candidates (e.g. $M_X \sim 10^{20}$ g), this condition is satisfied up to high redshift values ($z \sim 10^8$). Thus, we can safely assume that the dark matter fluid approximation is valid when dealing with DM-baryon interactions at the CMB epoch.}. 

Following \cite{McDermott:2010pa}, the momentum transfer rate due to DM-baryon scattering can be written as
\begin{equation}
\Gamma_\text{c} = \sum_{i = p,e} \frac{8 \sqrt{2 \pi} \alpha^2 q_X^2 n_i \mu_i^{1/2}}{3 M_X T^{3/2}} \ln \left( \frac{3 T \lambda_D}{|q_X| \alpha} \right) \, , \label{eq:Gamma_c}
\end{equation}
where 
\begin{equation}
\mu_i = \frac{M_X m_i}{M_X + m_i}
\end{equation}
is the DM-baryon reduced mass (which in the case of Macros reduces to the mass of protons/electrons, since $M_X \gg m_p \gg m_e$), $q_X$ is the charge of DM in units of the elementary charge, $\alpha \simeq 1/137$ is the fine structure constant and
\begin{equation}
\lambda_D = \sqrt{\frac{T}{4 \pi \alpha n_e}}
\end{equation}
is the Debye length of the cosmological plasma.

The condition for the tight coupling between baryons and DM is that the momentum transfer rate is much larger than the Hubble expansion rate,
$\Gamma_\text{c} \gg H $.
The tight coupling between baryons and charged DM has been largely studied in the context of millicharged DM, where the DM is made up of particles carrying an electric charge much smaller than the elementary charge (see e.g. \cite{Dubovsky:2003yn,Dolgov:2013una,McDermott:2010pa,Dubovsky:2001tr,Melchiorri:2007sq}). 
In these scenarios, from Eq. \eqref{eq:Gamma_c} one can easily see that $\Gamma_{\text{c}} / H \propto T^{-1/2}$ during the radiation era (neglecting the logarithmic dependence in the screening term). This means that the DM and baryons are not coupled at high redshifts and then eventually get coupled when $\Gamma_{\text{c}} / H$ becomes of order unity.

The scenario considered in this paper differs in one aspect: protons are being absorbed by Macros, thus their number density has a different scaling than the usual one (i.e. $n_b \propto T^3$). Hence, the number of target protons that enters in the momentum transfer rate \eqref{eq:Gamma_c} is a function of the Macro parameters and is obtained again by numerically integrating Eq. \eqref{eq:proton-sol}. 

The electric charge of Macros is instead fixed by the Macro surface potential and cross-section.
For Macros with surface potential $V_X$ and cross-section $\sigma_X$ expressed in units of eV and eV$^{-2}$ respectively, the charge is given by  
\begin{equation}
q_X = \frac{V_X (\sigma_X / \pi)^{1/2}}{\alpha} \, . \label{eq:q}
\end{equation}

\begin{figure}
	\centering
	\includegraphics[scale=0.5]{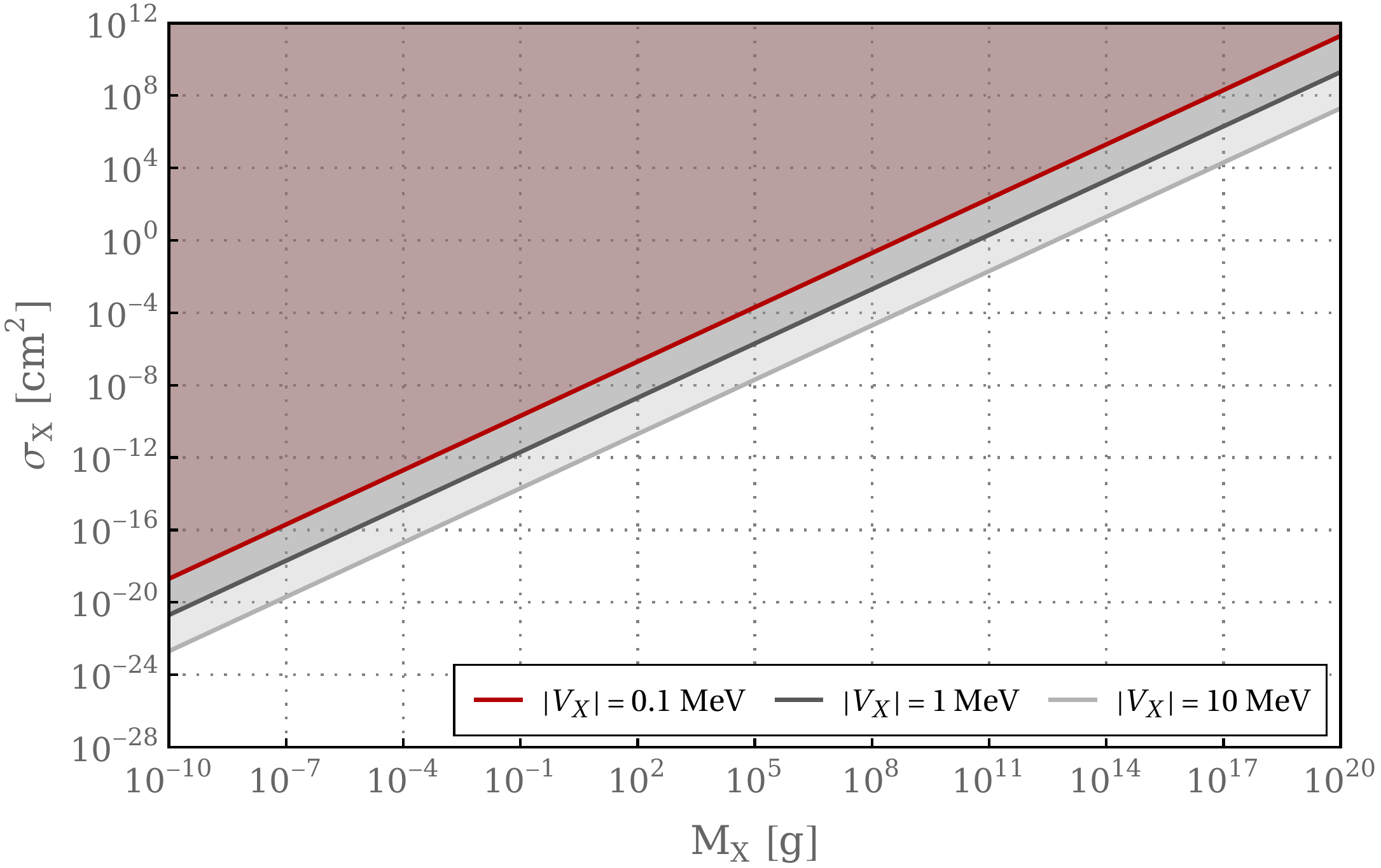}
	\caption{Constraints on the Macro cross-section and mass for different values of the surface potential $V_X$. These comes from the requirement that Macros and baryons are not tightly coupled due to Coulomb scattering at the recombination epoch.}
	\label{fig:tight-coupling}
\end{figure}

Now we have everything required to compute the momentum transfer rate given in Eq. \eqref{eq:Gamma_c}. CMB observations do not allow for a coupling between the two components at recombination, so we require $\Gamma_c/H < 1$ at that time. By imposing this condition, we derive the constraints shown in Figure \ref{fig:tight-coupling}. Note that these limits are likely conservative, because we expect that even relatively small momentum transfers between DM and baryons at the recombination epoch would leave a detectable imprint on CMB anisotropies. However, quantifying this effect would require a dedicated analysis, so for the time being we use the conservative condition $\Gamma_c/H < 1$.

Notice that the curves in Figure \ref{fig:tight-coupling} correspond to constant values of $\sigma_X / M_X$. Indeed, the momentum transfer rate $\Gamma_{\text{c}}$ depends on the ratio $q_X^2/M_X \propto \sigma_X/M_X$ and the proton number density also depends on $\sigma_X/M_X$ (see Eqs. \eqref{eq:rate-p}-\eqref{eq:proton-sol}). There is then the logarithmic dependence on $q_X \propto \sigma_X^{1/2}$, which is however very weak and gives negligible contributions. We can write the resulting constraints on the reduced cross-section as 
\begin{equation}
\frac{\sigma_X}{M_X} \lesssim 2 \cdot 10^{-11} \left( \frac{|V_X|}{\text{MeV}} \right)^{-2} \text{cm}^2 \, \text{g}^{-1} \, . \label{eq:constraints-tight-coupling}
\end{equation}

Before closing this section some comments are in order regarding our treatment of the charge of Macros and the relative velocity between MDM and baryons. Concerning the first aspect, we have assumed that the charge of Macros remains constant, despite the fact the protons are being captured. Since we are thinking of Macros as macroscopic nuclei, we might imagine that, as protons are absorbed, Macros get rid of the excess charge by converting protons to neutrons through weak processes.

\begin{figure}
	\centering
	\includegraphics[scale=0.5]{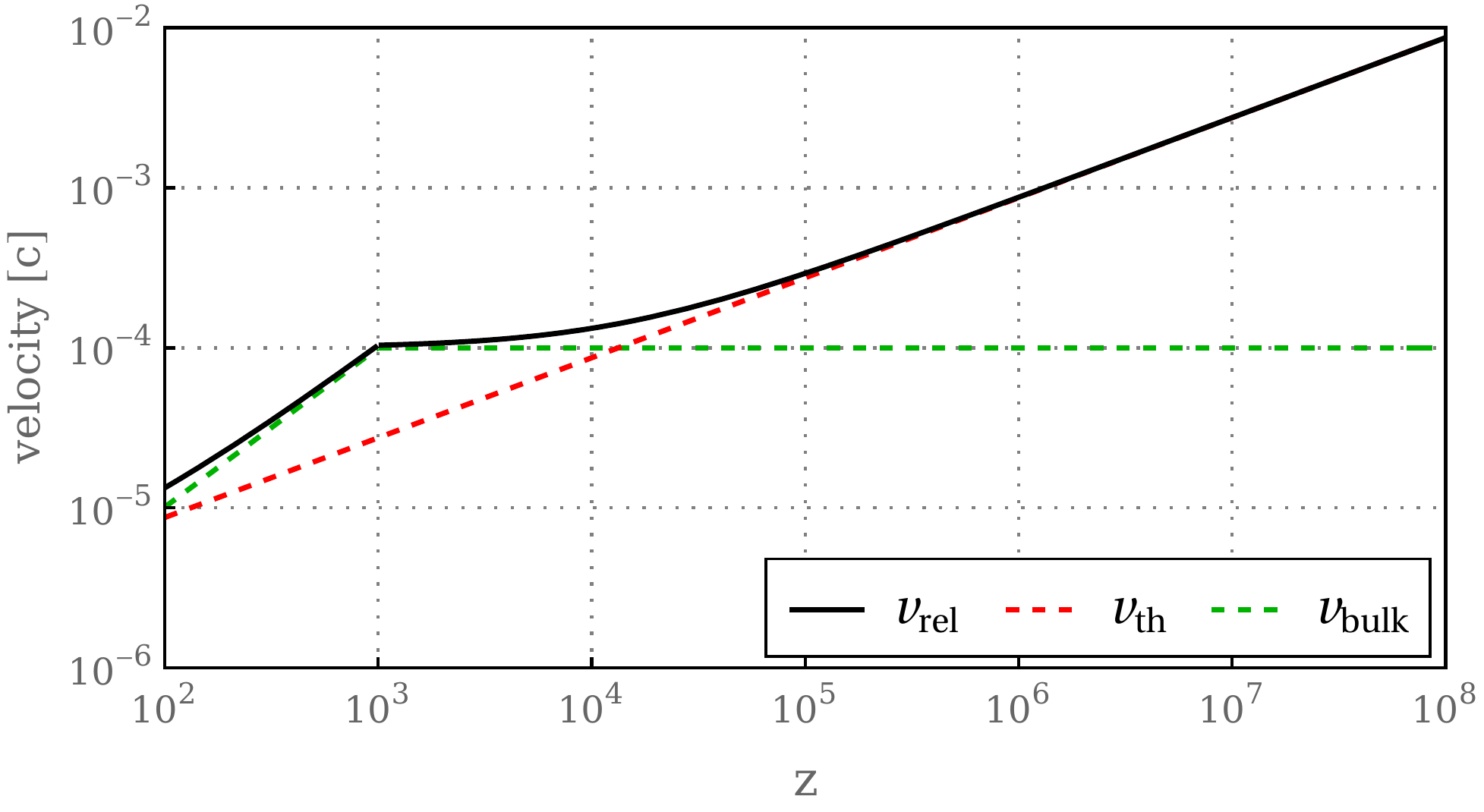}
	\caption{Relative bulk velocity between the baryon and DM fluids (red dashed line), thermal velocity of protons (blue dashed line) and relative velocity (black solid line) defined as $v_\text{rel} = (v_\text{th}^2 + v_\text{bulk}^2)^{1/2}$.}
	\label{fig:relative-velocity}
\end{figure}

As regards the relative velocity between baryons and Macros, this is in principle given by $v_\text{rel} = (v_\text{th}^2 + v_\text{bulk}^2)^{1/2}$, where $v_\text{th}$ is the thermal velocity of baryons and $v_\text{bulk}$ is the relative bulk velocity between the DM and baryon fluids. The latter is non-vanishing only if the two matter components are not coupled. If this is not the case, DM and baryons behave as a single fluid and the relative velocity is simply given by the thermal speed of baryons. A detailed treatment of the relative motion between the DM and baryon fluids can be found in \cite{Dvorkin:2013cea,Tseliakhovich:2010bj}. For the purposes of our discussion we can approximate the relative bulk velocity as 
\begin{equation}
v_\text{bulk}(z) \simeq \text{min} \left[ 1, \frac{1+z}{10^3} \right] \cdot 10^{-4} \, .
\end{equation}
This is constant at redshifts $z \gtrsim 10^3$ and then decreases as $(1+z)$. In Figure \ref{fig:relative-velocity} we show how the bulk, thermal and relative velocities evolve with redshift in the case in which DM and baryons are not coupled (i.e. $v_\text{bulk} \ne 0$).
It can be seen that at redshifts $z \gtrsim 10^4$ the thermal component is the dominant one and the approximation $v_\text{rel} \simeq v_\text{th}$ is then justified in this range of redshifts. Let us now discuss what happens at redshifts $z \lesssim 10^4$. First of all, notice that in this regime the bulk motion gives the largest contribution to $v_\mathrm{rel}$. It is then useful to recall that baryons and Macros are decoupled at high redshifts and get coupled only when $\Gamma_{\text{c}} / H$ becomes of order unity. If this condition is realized at redshifts $z \gtrsim 10^4$, then approximating the relative velocity with the thermal one is well justified throughout all the expansion history, since at $z \lesssim 10^4$ (i.e. when baryons and the DM are coupled) there is no relative bulk motion between the two components. If instead Macros and baryons become coupled at $z<10^4$ but before CMB decoupling, by assuming that $v_\text{rel} \simeq v_\text{th}$ in this range of redshifts we are underestimating the capture rate \eqref{eq:rate-p}, since $v_\text{th} < v_\text{rel}$. We have however verified that 
always approximating $v_\mathrm{rel}\simeq v_\mathrm{th}$ has a negligible impact on our results.

\subsection{Light element abundances}

Another possible effect due to photon injection is to alter the light element abundances after BBN \cite{Cyburt:2002uv,Ishida:2014wqa,Poulin:2015woa,Poulin:2015opa,Salvati:2016jng}. For the energies under consideration (i.e., $E_\gamma \lesssim 3.44$ MeV), two relevant processes to consider are the photodissociation of the deuterium and the $^7$Be, whose energy thresholds are given by $E_{\text{th}} = 2.2246$ MeV and $E_{\text{th}} = 1.5866$ MeV, respectively \cite{Cyburt:2002uv,Ishida:2014wqa}.
The equation which regulates nuclear abundances ($Y_A \equiv n_A/n_b$, with $A = d$, $^7$Be) in redshift space is given by (see e.g. \cite{Poulin:2015opa,Poulin:2015woa})
\begin{align}
\nonumber
\frac{dY_A}{dz} = -\frac{1}{H(z)(1+z)} &\left[ \sum_T Y_T \int_{0}^{\infty} dE_\gamma f_\gamma(E_\gamma,z)\sigma_{\gamma+T\rightarrow A}(E_\gamma) + \right. \\  
&\left. - Y_A \sum_P \int_{0}^{\infty} dE_\gamma f_\gamma(E_\gamma,z)\sigma_{\gamma+A\rightarrow P}(E_\gamma) \right] \, , \label{eq:abundances}
\end{align}
where $\sigma_{\gamma+T\rightarrow A}$ is the cross-section for the production of $A$ via the photodissociation of nuclei $T$, $\sigma_{\gamma+A\rightarrow P}$ is the cross-section for the analogous destruction channel and $f_{\gamma}$ is the nonthermal photon distribution function. The latter can be computed by solving the usual Boltzmann equation with the source term
\begin{equation}
S_{\gamma} = n_p(z) \Gamma_{p X}(z) p_\gamma \, ,
\end{equation}
where $p_\gamma$ is the injection spectrum. In the case of Macros composed of ordinary matter, the energy of the injected photons is given by the Macro binding energy, hence $p_\gamma = \delta (E_\gamma - I)$. As discussed in Section \ref{subsec:Thermodynamics}, the interaction rate of photons with the cosmological plasma, $\Gamma_{\gamma \text{Pl}}$, is much faster than the expansion rate $H$. As a result, $f_\gamma$ is driven to a quasi-static equilibrium (i.e. $\partial f_{\gamma}/\partial t =0$), such that $f_\gamma = S_\gamma / \Gamma_{\gamma \text{Pl}}$ \cite{Poulin:2015opa,Poulin:2015woa,Salvati:2016jng}.

For the energies under consideration, only destruction processes need to be considered. Thus, Eq. \eqref{eq:abundances} can be easily integrated as 
\begin{equation}
\ln \left( \frac{Y_A(z_f)}{Y_A(z_i)} \right) = \int_{z_i}^{z_f} \frac{dz}{H(z)(1+z)} \frac{n_p(z)\Gamma_{p X}(z)}{\Gamma_{\gamma \text{Pl}}(I,z)} \, \sigma_{\gamma+A\rightarrow P}(I) \, . \label{eq:abundances-fin}
\end{equation}
The explicit expressions of the photodissociation cross-sections of $d$ and $^7$Be can be found in Refs. \cite{Cyburt:2002uv} and \cite{Ishida:2014wqa}, respectively. 

It is now possible to evaluate the impact of the energy injection on the abundances of $d$ and $^7$Be by means of Eq. \eqref{eq:abundances-fin}.
We find that the resulting effects are negligible and do not allow us to constrain MDM. This happens because the emitted photons quickly thermalize with the baryon-photon plasma and are practically ``lost'' as photodissociation sources.

\section{Antimatter Macros}
\label{sec:anti-Macros}

In this section we consider the possibility that the dark matter is in the form of macroscopic objects composed of antimatter. Following \cite{SinghSidhu:2020cxw}, we refer to this class of DM candidates as anti-Macros. 
 
Before the recombination epoch, free protons of the cosmological fluid can be captured by anti-Macros, analogously to what happens in the case of Macros composed of ordinary matter. The annihilations between protons and antiprotons then lead to the release of high-energy photons in the baryon-photon plasma.

Differently to what happens in the case of Macros (see Eq. \eqref{eq:bcapture}), there is no inverse process available for keeping the comoving density of baryons constant at high redshifts. This implies that, as soon as anti-Macros are produced, they start absorbing protons. This also means that the constraints that we find are independent of the binding energy of anti-Macros. We leave instead the redshift at which anti-Macros are produced as a free parameter. To keep contact with the previous sections, we label again this as $z_\text{in}.$

Some regions of the parameter space of anti-Macros have already been constrained in \cite{SinghSidhu:2020cxw}, where the authors focused on the case with $V_X = 0$. Following the discussion of the previous sections, we now derive new constraints from the absorption of protons between the BBN and recombination and from spectral distortions of the CMB. Moreover, the condition that baryons and anti-Macros were not coupled at decoupling has still to be imposed. In Section \ref{sec:Tight-coupling} we have seen that the resulting constraints on the Macro parameter space are independent of the binding energy $I$. Since the latter fixes the redshift at which the capture of protons begins, this means that these constraints are independent of $z_\text{in}$. The same conclusion has to be true also in the case of anti-Macros, which then have to obey the same constraints as those shown in Figure \ref{fig:tight-coupling}.

\subsection{Baryon density between BBN and CMB epochs}
As we have done in Section \ref{subsec:baryon} for Macros, we can constrain the parameter space of anti-Macros by requiring that the number of protons that have been absorbed between the BBN and CMB epochs does not exceed the experimental constraints, i.e. $\Delta \log{\mathcal N}_p \le (\Delta \log{\mathcal N}_p)_\mathrm{obs}$ within the observational uncertainty.

Since the capture rate is equal to the case of Macros, these results coincide with those of Figure \ref{fig:const_baryon_numb}, with the only difference of $z_\text{in}$ replacing $I$ as a phenomenological free parameter. In particular, from Figure \ref{fig:const_baryon_numb} we can derive the constraints on anti-Macros through the following mapping:
\begin{align}
\nonumber
I &= 0.1 \, \text{MeV} \;\; \longrightarrow \; z_{\text{in}} \simeq 1.1 \cdot 10^7 \\
I &= 1 \, \text{MeV} \;\quad \longrightarrow \; z_{\text{in}} \simeq 1.2 \cdot 10^8 \label{eq:mapping-I-zin}\\
\nonumber
I &= 3.44 \, \text{MeV}  \longrightarrow \; z_{\text{in}} \simeq 4.3 \cdot 10^8 \, .
\end{align}

\subsection{CMB spectral distortions from $p \bar{p}$ annihilations}

If Macros are composed of antimatter, annihilations with protons of the cosmological plasma result in the release of high-energy photons. This process has been discussed in \cite{SinghSidhu:2020cxw}, where the reduced cross-section of anti-Macros has been constrained by analyzing the effects of this energy injection on the CMB anisotropies and the BBN. The heating rate depends on which specific process takes place after the $p\bar{p}$ annihilations (like, e.g., the nature of the cascade of particles that are produced). Denoting with $k$ the fraction of the rest energy of the proton and antiproton that is actually released in the baryon-photon plasma, we can write the heating rate as \cite{SinghSidhu:2020cxw}
\begin{equation}
\dot{Q} = n_p(z) \Gamma_{p X}(z)(2km_p) \, . \label{eq:Q-antiMacros}
\end{equation}
In the following we take $k = 0.2$. This is consistent with the scenario in which $p\bar{p}$ annihilations result in the production of multi-pion states, including both neutral and charged pions. The latter (or their decay products, i.e. electrons, muons and neutrinos) are able to escape anti-Macros, while neutral pions decay after their production into four photons, each having energy $E_\gamma \sim 100$ MeV \cite{SinghSidhu:2020cxw}. 

\begin{figure}
	\centering
	\includegraphics[scale=0.66]{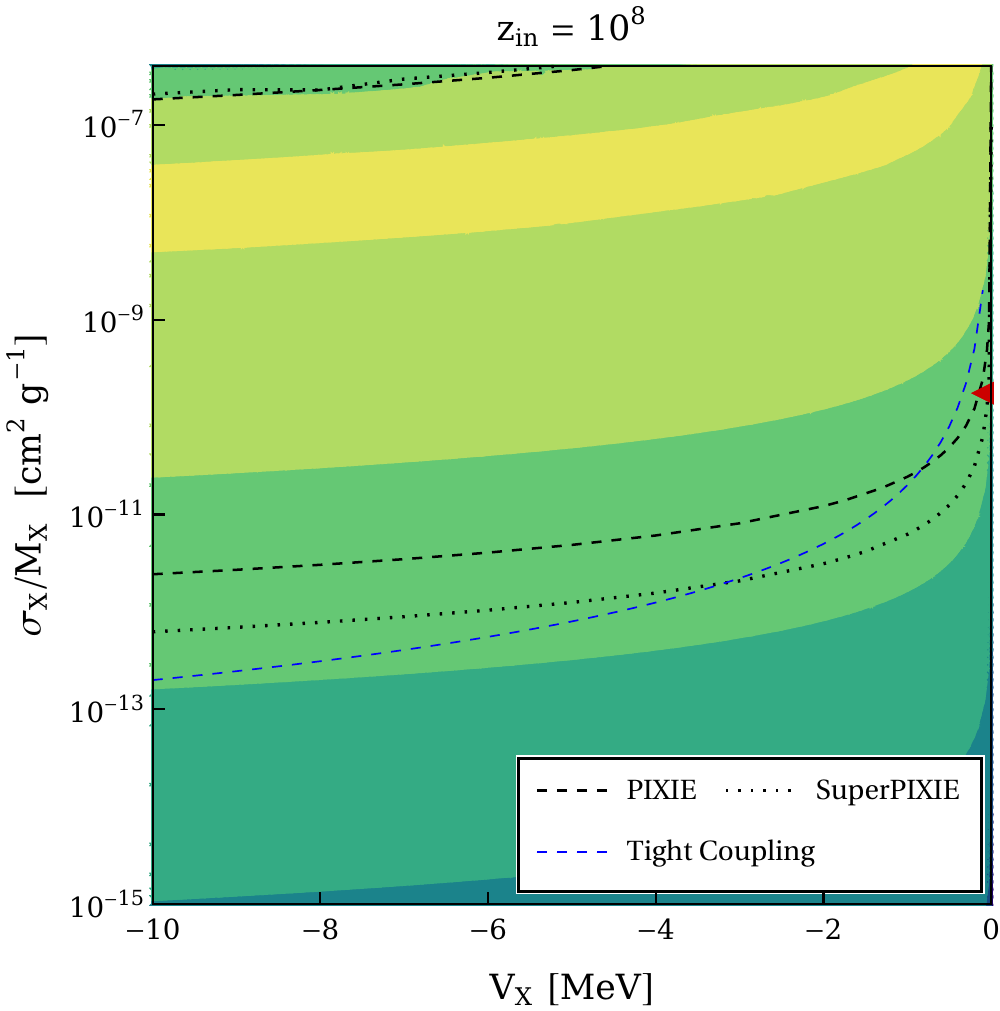}
	\includegraphics[scale=0.66]{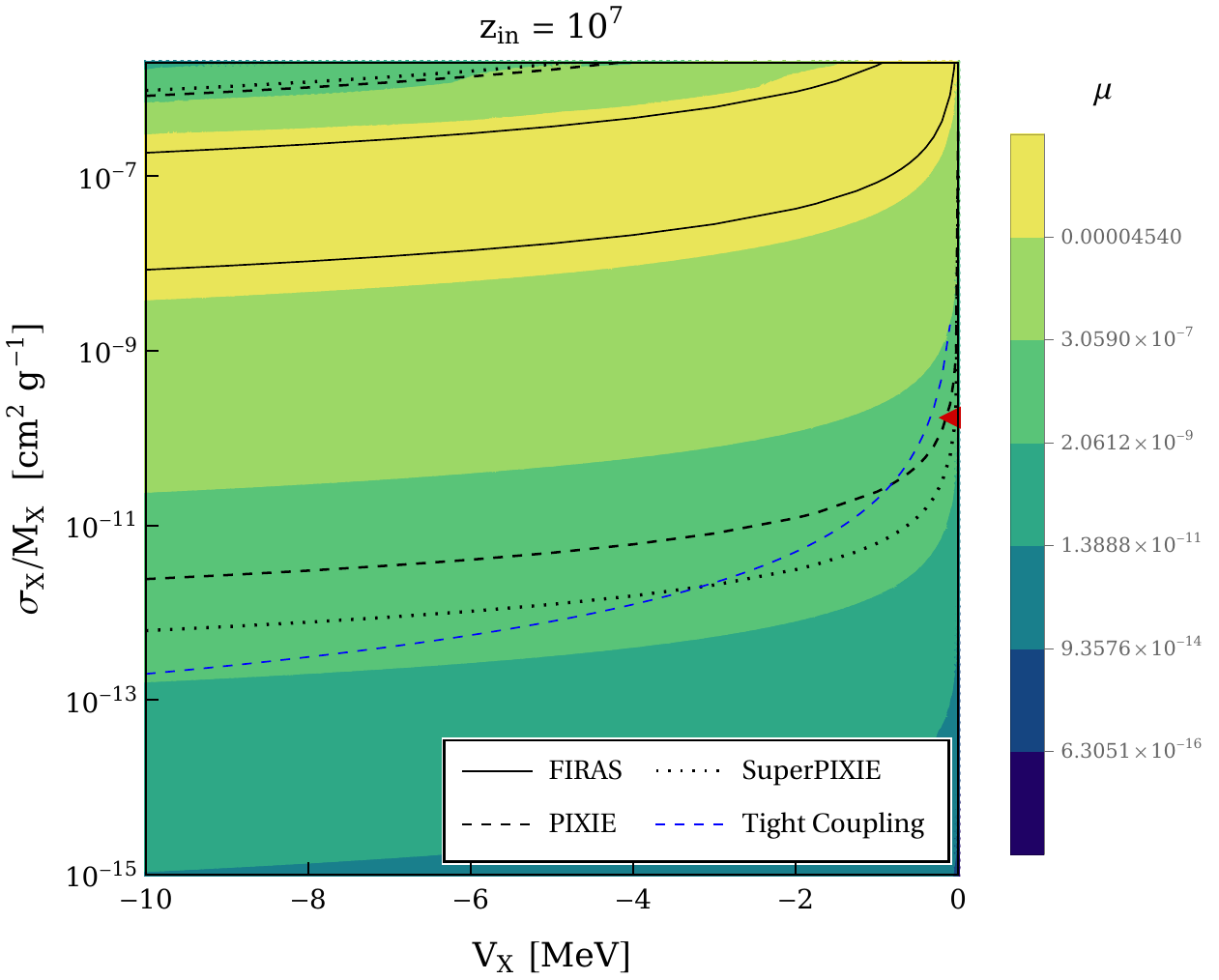}\vspace{2mm}
	\includegraphics[scale=0.66]{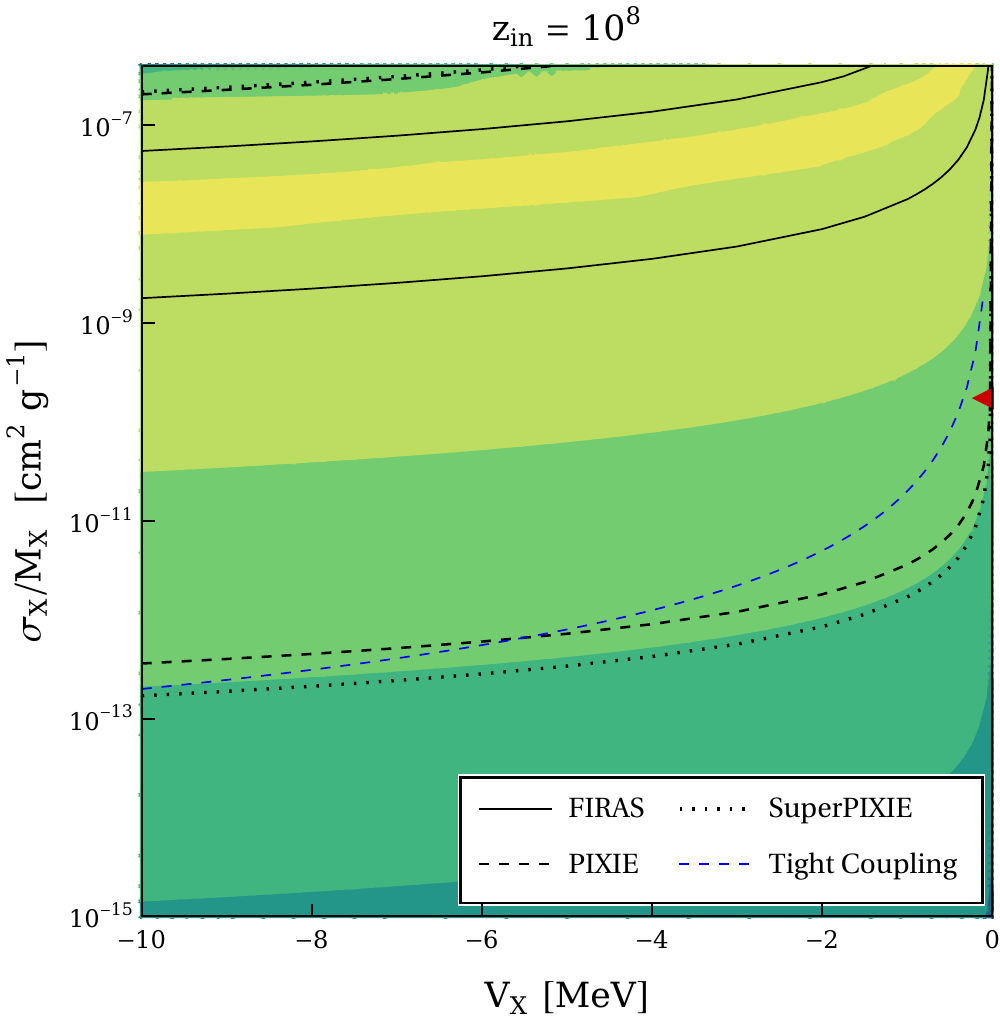}
	\includegraphics[scale=0.66]{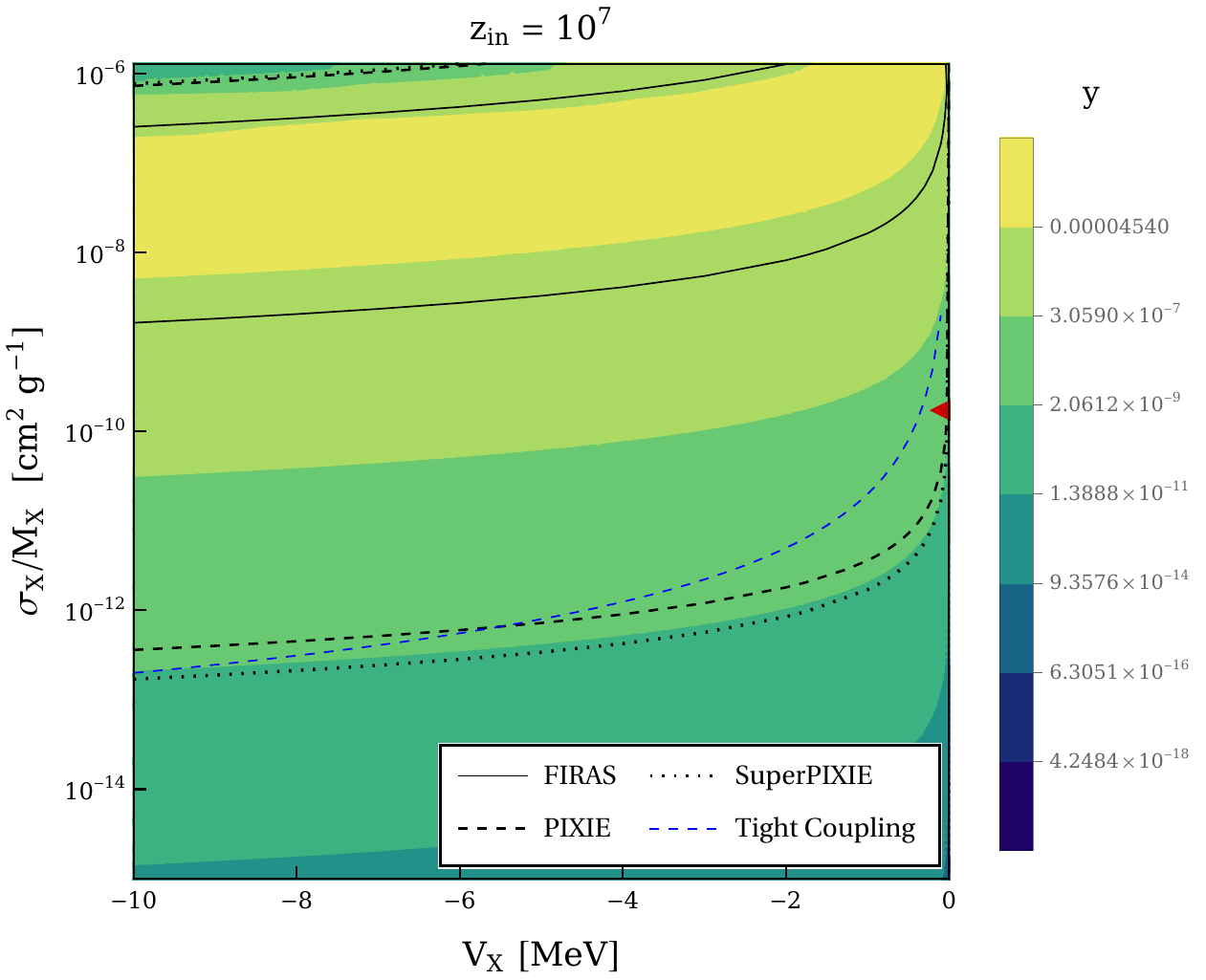}
	\caption{Contour plots of $\mu$ and $y$-distortions as a function of the reduced cross-section $\sigma_X / M_X$ and the surface potential $V_X$ of anti-Macros, for two different values of the initial redshift: $10^7$ and $10^8$. These are due to the emission of high energy photons as a result of $p\bar{p}$ annihilations. The region of parameter space within the black solid lines is excluded by FIRAS ($|\mu| < 9 \cdot 10^{-5}$, $|y| < 1.5 \cdot 10^{-5}$). The regions of parameter space within the black dashed and dotted lines are excluded assuming a null detection of $\mu$-distortions from PIXIE ($|\mu| < 3 \cdot 10^{-8}$, $|y| < 3.4 \cdot 10^{-9}$) and SuperPIXIE ($|\mu| < 7.7 \cdot 10^{-9}$, $|y| < 1.6 \cdot 10^{-9}$), respectively. The dot-dashed blue curves correspond to the bounds \eqref{eq:constraints-tight-coupling}, obtained by requiring that anti-Marcos and baryons are not coupled at decoupling. The red point denotes the constraint derived in \cite{SinghSidhu:2020cxw}, which has been obtained for $V_X = 0$.}.
	\label{fig:distortions-AntiMacro}
\end{figure}

The amount of $\mu$ and $y$-distortions produced by anti-Macros can then be computed through Eqs. \eqref{eq:mu-distorsions}-\eqref{eq:y-distorsions}. 
The results are shown in Figure \ref{fig:distortions-AntiMacro} for two different values of the initial redshift, $z_\text{in} = 10^8$ and $z_\text{in} = 10^7$. We also show the constraints obtained exploiting the upper bounds on spectral distortions by FIRAS, as well as the forecasted constraints obtained assuming a null detection of spectral distortions by PIXIE and SuperPIXIE. The constraints derived in \cite{SinghSidhu:2020cxw} for $V_X = 0$ are also included and are marked with a red triangle. 

Notice in particular that in this case it is possible to exclude some regions of parameter space thanks to the upper bounds on spectral distortions set by FIRAS. However, these bounds are weaker than those obtained by the capture of protons and the tight coupling condition, as we will also discuss in the next section. On the other hand, PIXIE and SuperPIXIE would allow us to improve the current constraints. 

\section{Discussion and conclusions}
\label{sec:Conclusions} 

In this paper we have considered the possibility that dark matter consists of macroscopic-size objects, focusing in particular on the cosmological phenomenology associated with these DM candidates. 
Without referring to any specific model, we have analyzed two scenarios in which Macros are composed of either ordinary matter or antimatter. In both cases, Macros can absorb protons from the cosmological plasma between the BBN and the CMB decoupling. This results in the release of high energy photons in the baryon-photon fluid.
We derive constraints on the Macro parameter space from three cosmological processes: the variation of the baryon density between the BBN and the CMB epochs, the tight coupling between charged Macros and baryons at recombination and the production of CMB spectral distortions. For the latter, we have also explored the capability of the proposed spectral distortions experiments PIXIE and SuperPIXIE. We have also analyzed how the energy injection affects the abundance of light elements after the BBN. In this regard, we find that the effects are negligible and do not allow us to further constrain the Macro parameter space. 

\begin{figure}
	\centering
	\includegraphics[scale=0.355]{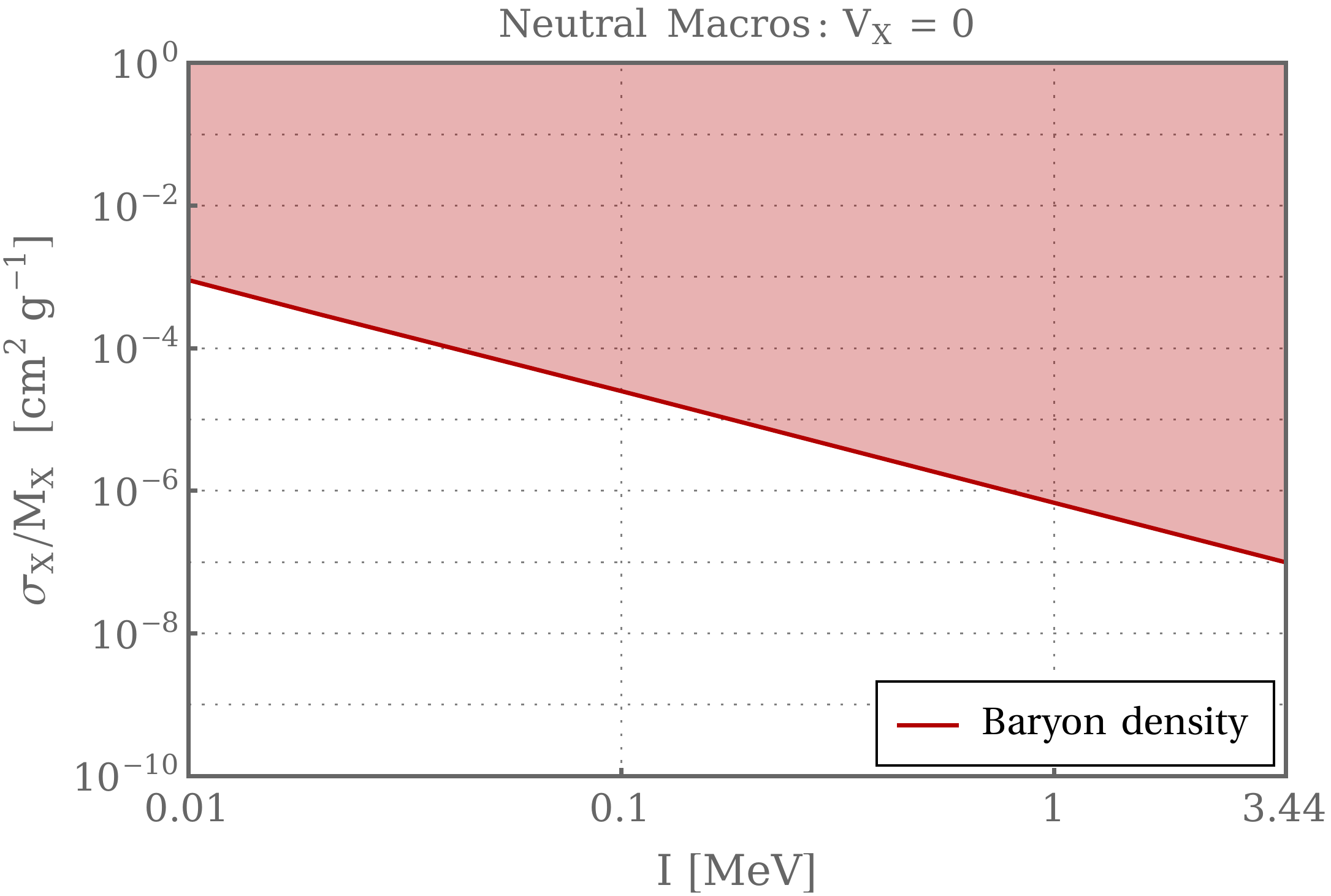}
	\includegraphics[scale=0.342]{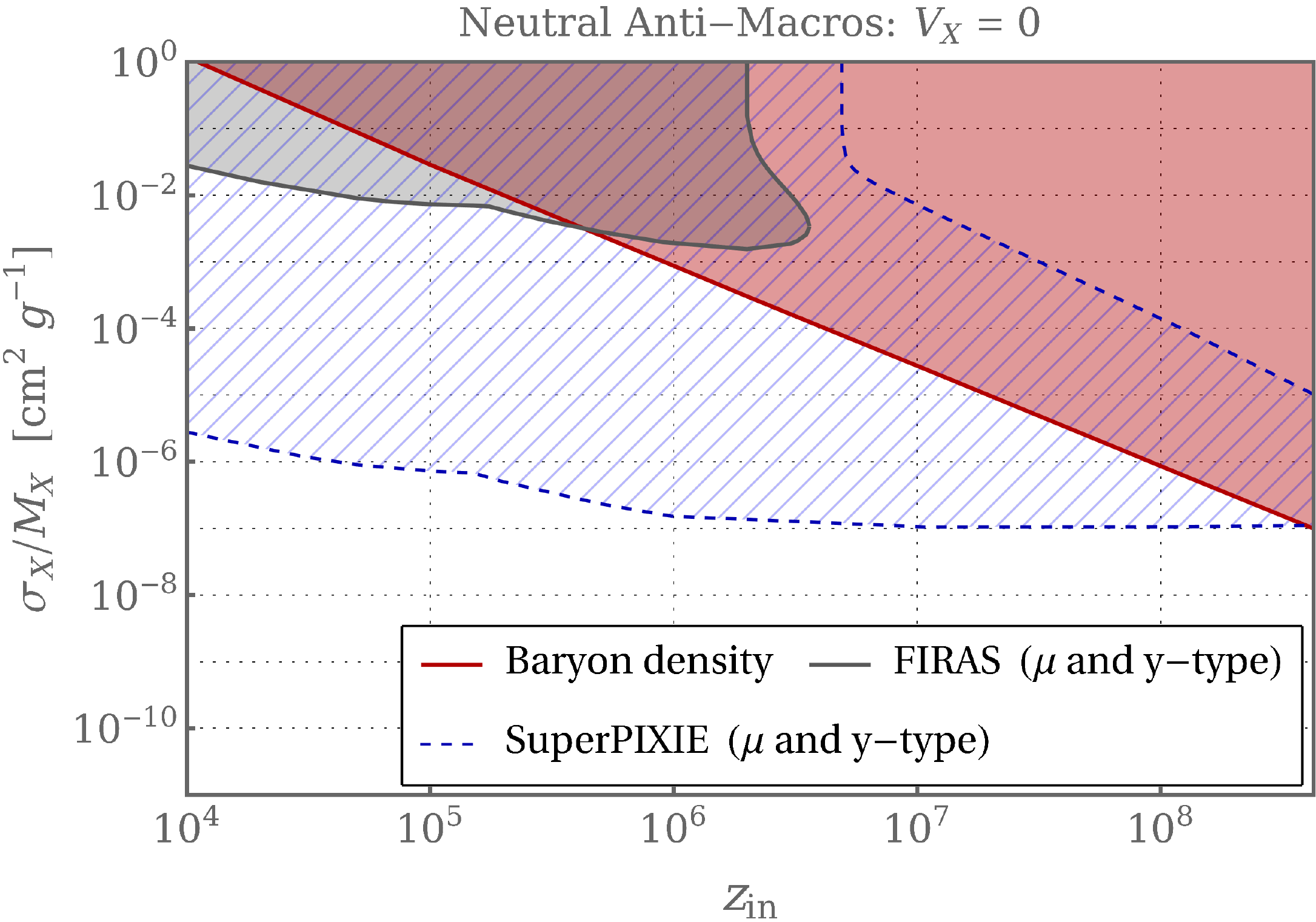}\vspace{2mm}
	\includegraphics[scale=0.355]{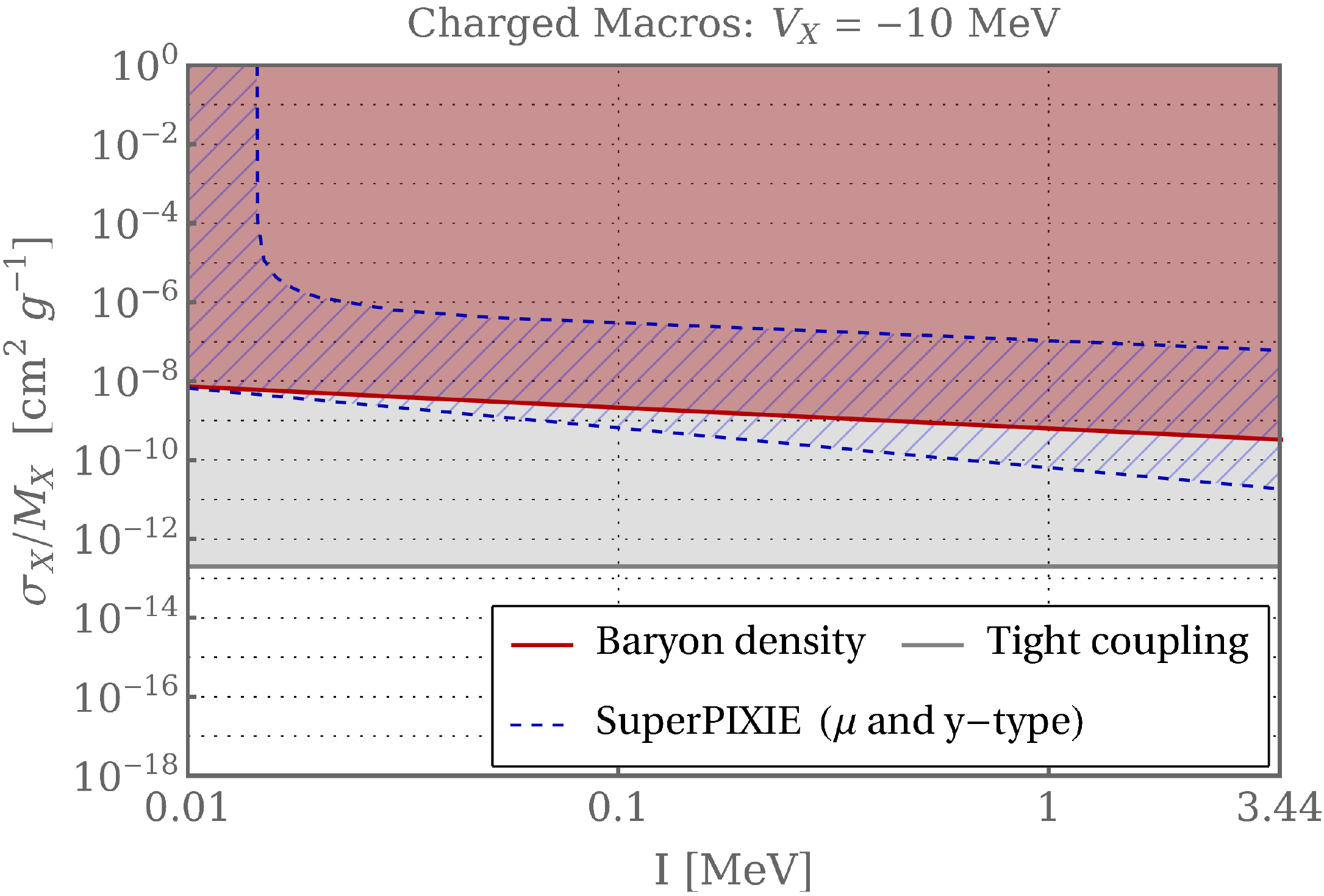}
	\includegraphics[scale=0.355]{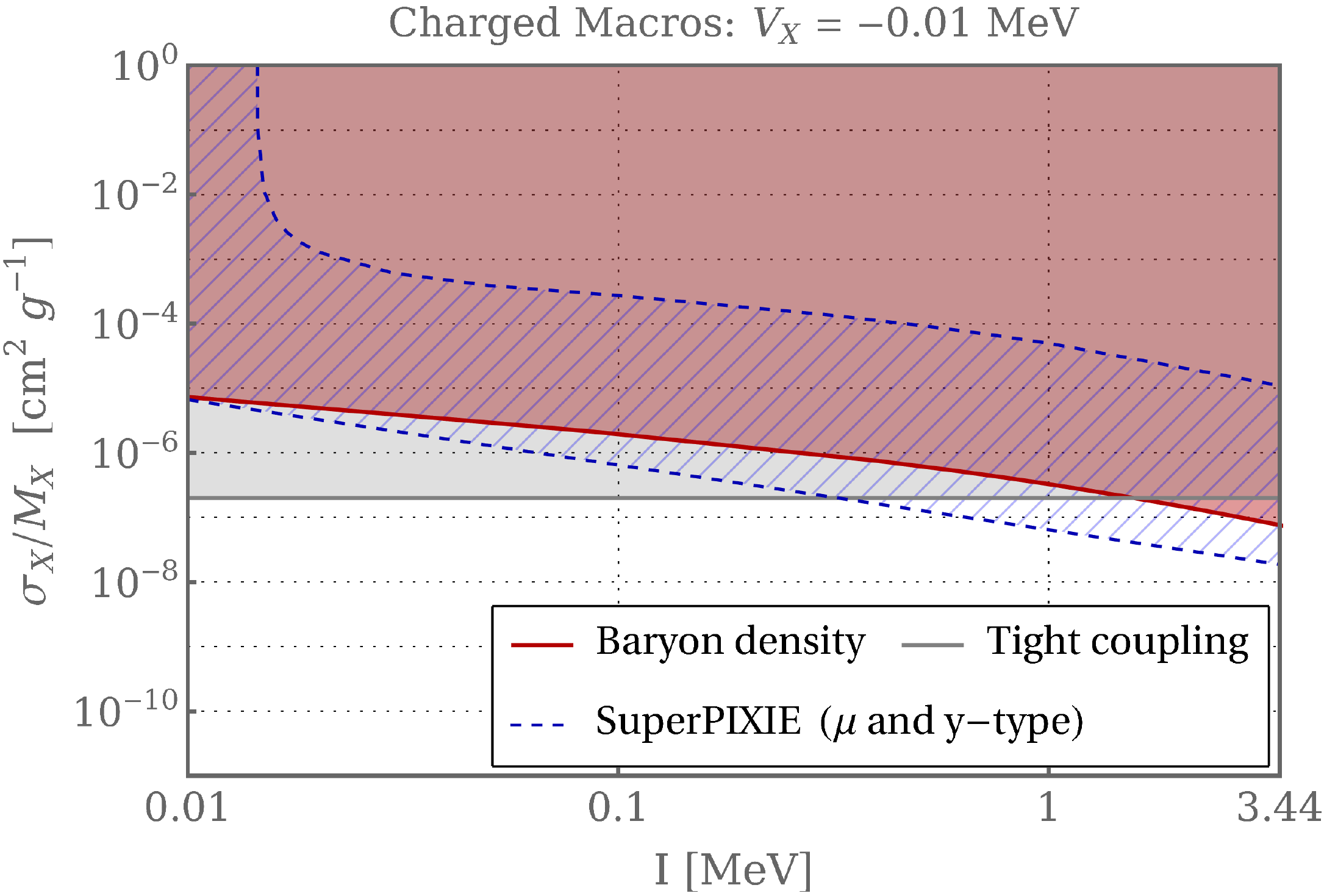}\vspace{2mm}
	\includegraphics[scale=0.345]{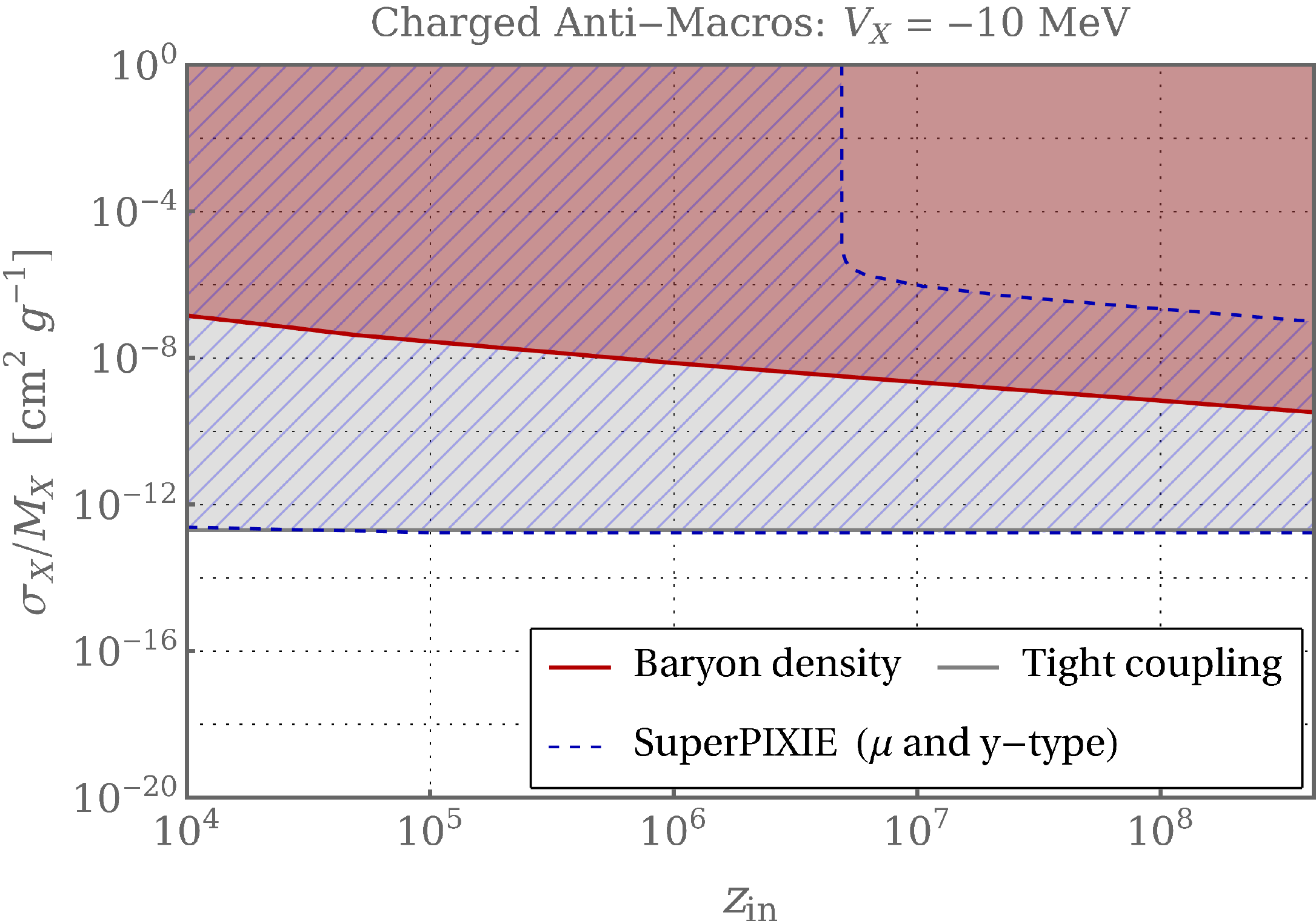}\hspace{1mm}
	\includegraphics[scale=0.345]{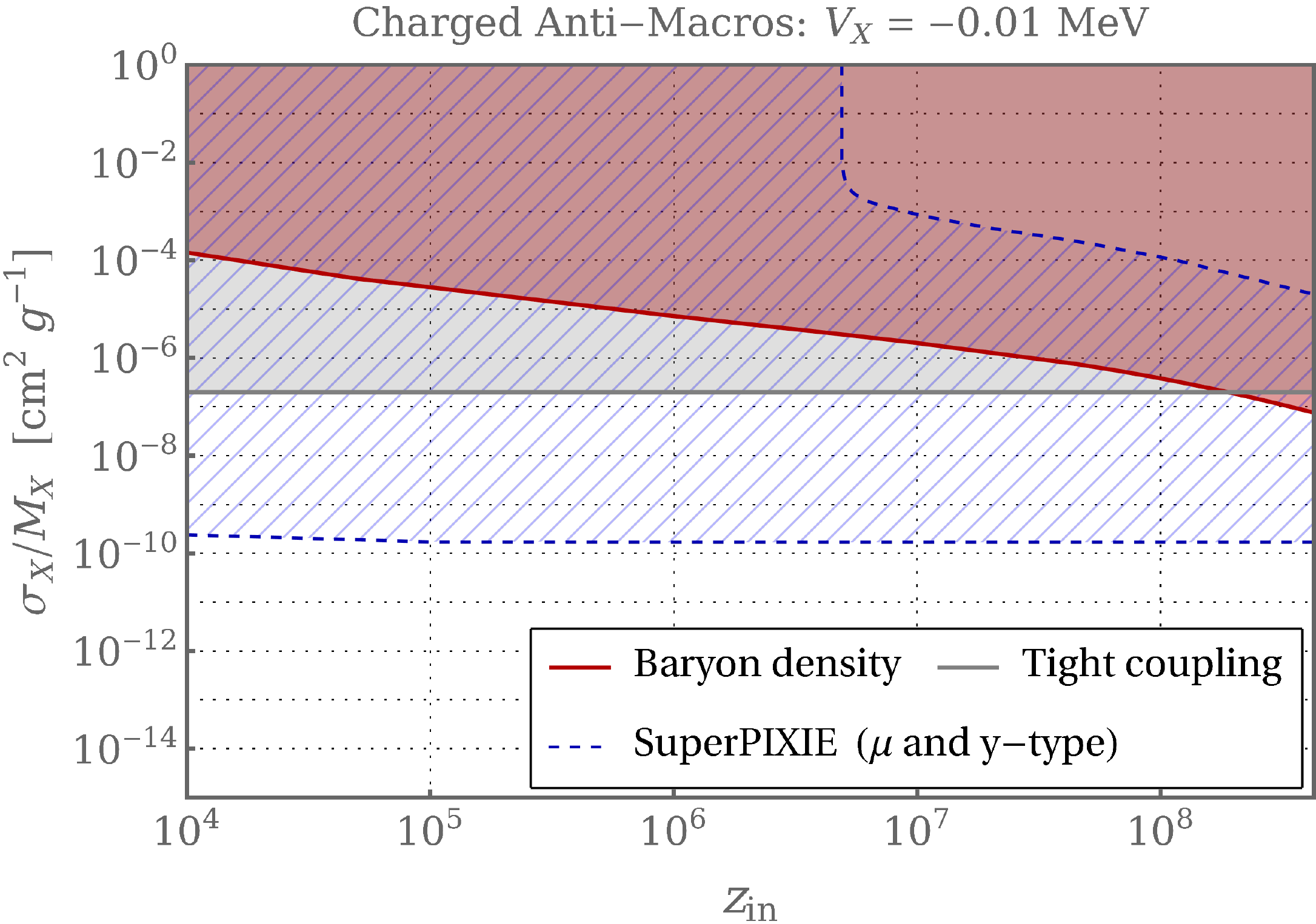}
	\caption{Constraints on the reduced cross-section as a function of the Macro binding energy $I$ (or the initial redshift $z_\text{in}$, in the case of anti-Macros), for different values of the surface potential $V_X$. The shaded regions are excluded by current constraints, while the hatched regions represent the sensitivity windows for SuperPIXIE. In the upper panels we show the constraints on neutral Macros (upper left) and neutral anti-Macros (upper right). In the central panels, instead, the constraints on charged Macros are reported for two different values of the surface potential, $V_X = -10$ MeV (central left) and $V_X = -0.01$ MeV (central right). Analogous bounds for charged anti-Macros are shown in the lower panels, for $V_X = -10$ MeV (bottom left) and $V_X = -0.01$ MeV (bottom right). The constraints from the baryon density and from spectral distortions are valid for $\sigma_X/M_X \lesssim 1 \, \mathrm{cm}^2 \, \mathrm{g}^{-1}$, since for higher values of the reduced cross-section the injected photons are not absorbed efficiently by the cosmological plasma. The bounds derived from the tight coupling condition are instead valid also for $\sigma_X/M_X > 1 \, \mathrm{cm}^2 \, \mathrm{g}^{-1}$.}
	\label{fig:summary}
\end{figure}

The constraints that we obtain are summarized in Figure \ref{fig:summary}. In the following we will comment on the main features of these results.
Let us start by discussing the case of neutral MDM: 
\begin{itemize}
	\item For Macros composed of ordinary matter, the only constraints we find are derived by imposing that the amount of protons absorbed between the BBN and the CMB epochs does not exceed the observational bound \eqref{eq:Delta_Np}. The resulting constraints are shown in the top left panel of Figure \ref{fig:summary}. An analytical fit to these is given by 
	\begin{equation}
	\frac{\sigma_X}{M_X} \lesssim 6.8 \cdot 10^{-7} \left( \frac{I}{\text{MeV}} \right)^{-1.56} \, \text{cm}^2 \, \text{g}^{-1} \, . \label{eq:constraints-neutral}
	\end{equation}
	No sizeable spectral distortions are produced, implying that future spectral distortions experiments, like PIXIE and SuperPIXIE, would not allow to improve the bound \eqref{eq:constraints-neutral}.
	
	\item For neutral anti-Macros, the current constraints are dominated by the baryon density condition for $z_\text{in} \gtrsim 4 \cdot 10^5$, while for $z_\text{in} \lesssim 4 \cdot 10^5$ the FIRAS bounds on spectral distortions give the tightest constraints on the reduced cross-section. 
	Future CMB spectral distortion experiments will probe a much larger region of the parameter space, as can be appreciated in the top right panel of Figure \ref{fig:summary}.
\end{itemize}

In the case of charged MDM with strong enough surface potential, the tightest constraints come from requiring that Macros and baryons are not coupled at recombination due to Coulomb scattering\footnote{Notice that in Figure \ref{fig:summary} we do not report the constraints on charged anti-Macros derived from the FIRAS bounds on spectral distortions. Indeed, as can be seen in Figure \ref{fig:distortions-AntiMacro}, these are always below the constraints derived from the tight coupling condition.} ($\Gamma_\text{c} / H < 1$). For vanishingly small potentials (see e.g. the $V_X \simeq -0.01$ MeV case shown in the center right panel of Figure \ref{fig:summary}) the bounds from the baryon density can be tighter in some region of parameter space.

The tight coupling condition leads to the following conservative bound on the reduced cross-section (see also Figure \ref{fig:tight-coupling})
\begin{equation}
\frac{\sigma_X}{M_X} \lesssim 2 \cdot 10^{-11} \left( \frac{|V_X|}{\text{MeV}} \right)^{-2} \text{cm}^2 \, \text{g}^{-1} \, , \label{eq:constraints-tight-coupling-2}
\end{equation}
where $V_X$ denotes the Macro surface potential.
An interesting point to remark is that these constraints are basically insensitive to both the Macro binding energy (or $z_\text{in}$, in the case of anti-Macros) and the sign of the surface potential. Our interpretation for this goes as follows: first of all, the sign of the surface potential $V_X$ determines the sign of the Macro electric charge through Eq. \eqref{eq:q}, but this does not affect the momentum transfer rate \eqref{eq:Gamma_c}, which depends only on the absolute value of the charge. Secondly, both $I$ and $V_X$ affect the behavior of $n_p(z)$, but $\Gamma_c$ does depend on the combination $n_p \mu_p + n_e \mu_e$. Since we are assuming that electrons are not absorbed by Macros, even if a non-negligible fraction of protons is absorbed, the value of the momentum transfer rate does not change too much, given that all the electrons are still available as scattering targets.

We stress that, because of this last point, the constraints from the tight coupling condition are valid also in the region of parameter where $\sigma_X/M_X \gtrsim \, 1 \text{cm}^2 \, \text{g}^{-1}$, contrary to constraints from the baryon density and spectral distortions. Indeed, as discussed above, the constraints from the tight coupling are basically only due to the charge of Macros, which is not related to the absorption of protons, but rather remains constant at its initial value. Therefore, Coulomb scattering takes place also in the region of parameter space with $\sigma_X/M_X \gtrsim \, 1 \text{cm}^2 \, \text{g}^{-1}$ and the condition \eqref{eq:constraints-tight-coupling-2} has still to be satisfied.

It is then important to stress how future spectral distortions experiments would allow to improve these constraints in the case with $V_X < 0$:
\begin{itemize}
	\item For Macros composed of ordinary matter, we find that future spectral distortions experiments would improve the bounds on the reduced cross-section for small values of $|V_X| \lesssim 0.05$ MeV. This can be seen in the middle right panel of Figure \ref{fig:summary}, where we report the sensitivity region for SuperPIXIE including both $\mu$ and $y$-type spectral distortions for the case with $V_X = -0.01$ MeV.
	
	\item For anti-Macros, we can see that the sensitivity window for SuperPIXIE roughly coincides with the region of parameter space excluded by the the tight coupling condition for $V_X = -10$ MeV, see bottom left panel of Figure \ref{fig:summary} . For smaller values of the surface potential, SuperPIXIE would instead improve the bounds on anti-Macros. This is shown in the lower right panel of Figure \ref{fig:summary} for the case with $V_X = -0.01$ MeV.   
\end{itemize} 

We now want to compare our results with the cosmological constraints on Macros derived in previous literature.
In particular, some constraints on $\sigma_X/M_X$ as a function of $V_X$ have been derived in Ref. \cite{Jacobs:2014yca}. However, there are some key differences with respect to our analysis that we want to outline. The constraints obtained in \cite{Jacobs:2014yca} come from the requirement that the standard BBN predictions are not altered by the interactions between Macros and baryons (both protons and neutrons). Indeed, phrased within our theoretical framework, the authors of Ref.~\cite{Jacobs:2014yca} consider the case $I > I_{\text{BBN}}$, since the absorption of baryons starts before BBN. 
This means that we are actually probing different regions of the parameter space.
It is also worth mentioning that the constraints derived in \cite{Jacobs:2014yca} do not take into account the injection of photons in the primordial plasma, but are uniquely based on protons and neutrons being captured with a different rate by Macros. 
Since the absorption of baryons occurs earlier in their scenario, a non negligible fraction of baryons is absorbed also for $V_X > 0$, provided that $V_X \sim \mathcal{O}(\text{MeV})$. 
In our case, instead, the lower thermal energy of protons makes it difficult to overcome a potential barrier $V_X \gtrsim 1$ MeV. In particular, only if $I \sim I_{\text{BBN}}$, i.e. if the absorption of protons starts just after the BBN, we find competitive constraints on the reduced cross-section for $V_X \sim 1$ MeV.
On the other hand, the BBN constraints derived in \cite{Jacobs:2014yca} are insensitive to the case with $V_X = 0$, when protons and neutrons are absorbed with the same rate, thus not affecting primordial abundances. Baryon absorption between BBN and decoupling, instead, is also sensitive to the case with $V_X = 0$, as we have already seen. 

Energy injection resulting from $p\bar{p}$ annihilations by anti-Macros has also been considered in Ref. \cite{SinghSidhu:2020cxw}. There, the authors derived a bound on $\sigma_X/M_X$ from the effects of the energy injection on CMB anisotropies and BBN, which results in $\sigma_X/M_X \lesssim 2 \cdot 10^{-10} \, \text{cm}^2\,\text{g}^{-1}$. Differently from our analysis, this bound has been derived only in the case with $V_X = 0$.   

It is interesting to briefly comment on the implications of our findings for Macro particles with a density of the order of the one of nuclear matter, $\rho_\mathrm{nuc} = 3.6 \times 10^{14} \,\mathrm{g}\,\cm^{-3}$. It is straightforward to see that an upper limit on the reduced cross section $\sigma_X/M_X < \alpha$ will in this case translate to a lower limit on the mass $M_X/\mathrm{g} > 1.4\times 10^{-29} \alpha^{-3}$. In the case of 
neutral macros, this yields  $M_X > 1.4 \times 10^{-8}\,\mathrm{g}$ for a binding energy $I=3.44 \,\MeV$, coming from the requirement that 
baryons are not absorbed in large amounts between the times of BBN and CMB decoupling. For charged macros with negative surface potential, the tight-coupling constraint in Eq. (\ref{eq:constraints-tight-coupling-2}) implies $M_X > 1.7 \times 10^{3}\,\mathrm{g}\,(|V_X|/\MeV)^6$.

\acknowledgments

We warmly thank Alessandro Gruppuso for suggesting the idea to look at the absorption of baryons by dark matter.
We thank Alessandro Gruppuso, Paolo Natoli, and Luca Pagano for frequent discussions while the project was being developed, and for useful comments on our manuscript. We also thank Alessandro Drago and Giuseppe Pagliara for useful discussion about strange matter.
We acknowledge support from the COSMOS network (www.cosmosnet.it) through the ASI (Italian Space Agency) Grants 2016-24-H.0, 2016-24-H.1-2018 and 2019-9-HH.0.

\newpage
\appendix

\section{Interaction rates between injected photons and the baryon-photon plasma}
\label{App:int-rates}

In this section we report the explicit expressions of the interaction rates between the injected photons and the cosmological plasma, following the results obtained in Refs. \cite{Zdziarski1989AbsorptionOX,Kawasaki:1994sc} (see also \cite{Salvati:2016jng}). As written in Eq. \eqref{eq:intgplasma}, the total interaction rate is given by the sum of different contributions: Compton scattering, photon scattering, pair production over nuclei and pair production over photons. These are functions of the redshift and the energy of the injected photons.

The interaction rate for Compton scattering with the electrons of the baryon-photon plasma can be written as \cite{Zdziarski1989AbsorptionOX}

\begin{equation}
\Gamma_{\text{Comp}} = n_b \, \sigma_T \left( \frac{1+2f_{\text{He}}}{1+4f_{\text{He}}} \right) (1+z)^3 f(x) \, ,
\end{equation} 
where $\sigma_T$ is the Thomson cross-section and we have defined
\begin{equation}
f(x) = \frac{3}{8x} \left[ \left(1-\frac{2}{x}-\frac{2}{x^2}\right) \ln (1+2x) + \frac{4}{x} +\frac{2x(1+x)}{(1+2x)^2} \right] \, ,
\end{equation}
with $x \equiv E_\gamma / m_e$. 

The interaction rate for photon scattering has instead the following expression \cite{Zdziarski1989AbsorptionOX}
\begin{equation}
\Gamma_{\text{PS}} = \frac{4448 \pi^4}{455625} \, \alpha^4 \lambda_c^{-1} \left( \frac{T_0}{m_e} \right)^6 x^3 (1+z)^6 \, ,
\end{equation}
where $\lambda_c = 1 / m_e$ is the Compton wavelength, $\alpha \simeq 1/137$ is the fine structure constant and $T_0 \simeq 2.35 \cdot 10^{-4}$ eV denotes the present-day CMB temperature.

Considering interactions of photons with both nuclei of H and $^4$He, the rate for pair production over nuclei can be written as \cite{Kawasaki:1994sc}
\begin{equation}
\Gamma_{\text{PPn}} = 
	\begin{cases}
	n_b \left[ \sigma_{x < 4}^{\text{H}}(x) \left(\frac{1+2f_{\text{He}}}{1+4f_{\text{He}}}\right) + \sigma_{x < 4}^{\text{He}}(x) \left(\frac{f_{\text{He}}}{1+4f_{\text{He}}}\right) \right] (1+z)^3 \quad \text{if} \;  x < 4 \vspace{3mm} \\ 
	n_b \left[ \sigma_{x \ge 4}^{\text{H}}(x) \left(\frac{1+2f_{\text{He}}}{1+4f_{\text{He}}}\right) + \sigma_{x \ge 4}^{\text{He}}(x) \left(\frac{f_{\text{He}}}{1+4f_{\text{He}}}\right) \right] (1+z)^3 \quad \text{if} \;  x \ge 4 \, ,
	\end{cases} 
\end{equation}
where
\begin{align}
\sigma_{x < 4}^{\text{A}} &= \frac{2\pi}{3} Z^2 \alpha r_e^2 \left( \frac{x-2}{x} \right)^3 \left[ 1 + \frac{1}{2}\rho + \frac{23}{40}\rho^2 + \frac{11}{60}\rho^3 + \frac{29}{960}\rho^4 + \mathcal{O}(\rho^5) \right] \, , \\
\nonumber
\sigma_{x \ge 4}^{\text{A}} &= 	Z^2 \alpha r_e^2 \bigg\{ \frac{28}{9}\ln 2x - \frac{218}{27} + \left( \frac{2}{x} \right)^3 \left[ \frac{2}{3}\left( \ln 2x \right)^3 - \left( \ln 2x \right)^2 + \left( \frac{6-\pi^2}{3} \right)\ln 2x + 2\zeta(3) + \right. \\ 
& \qquad\qquad\;\, \left. + \frac{\pi^2}{6} - \frac{7}{2} \right] - \left( \frac{2}{x} \right)^4 \left( \frac{3}{16}\ln x + \frac{1}{8} \right) - \left( \frac{2}{x} \right)^6 \left( \frac{29}{2304}\ln x + \frac{77}{13824} \right) + \mathcal{O}(x^{-8}) \bigg\} \, ,
\end{align}
where Z is the charge of the A nucleus, $r_e = \alpha / m_e$ is the classical radius of the electron and we have also defined
\begin{equation}
\rho \equiv \frac{2x-4}{x+2+2\sqrt{2x}} \, .
\end{equation}

Finally, the interaction rate for pair production over photons is given by \cite{Zdziarski1989AbsorptionOX}
\begin{equation}
\Gamma_{\text{PP}\gamma} = n_b \, \sigma_T \left( \frac{1+2f_{\text{He}}}{1+4f_{\text{He}}} \right) \left[ \frac{2\sqrt{\pi}}{\sqrt{y}e^{1/y}} \left(1+\frac{9}{4}y\right) \right] (1+z)^3 \, ,
\end{equation}
where we have defined
\begin{equation}
y \equiv x \frac{T_0}{m_e} (1+z) \, .
\end{equation}

\begin{figure}
	\centering
	\includegraphics[scale=0.45]{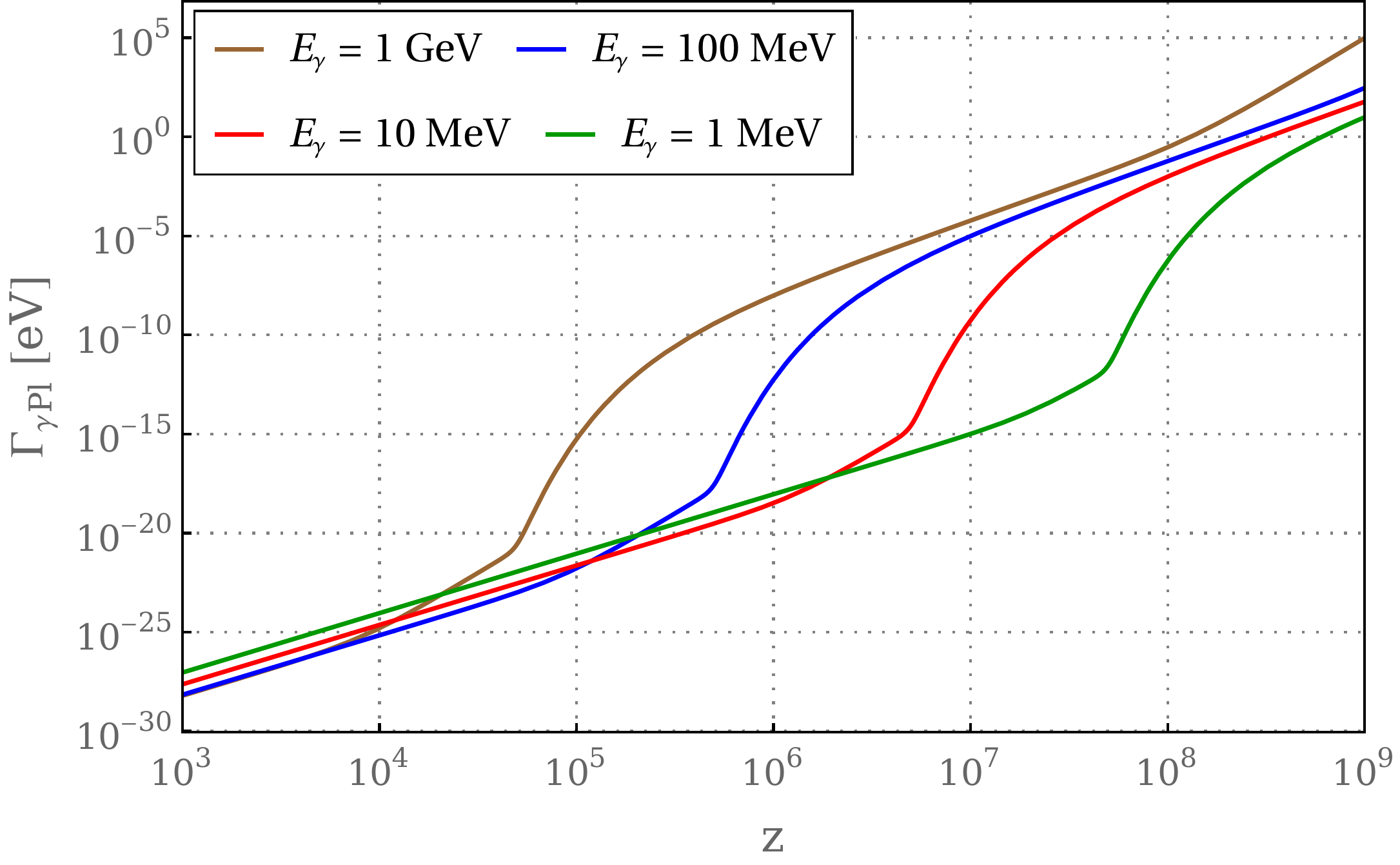}
	\caption{Total interaction rate for different values of the energy of the emitted photons: 1 MeV (brown), 10 MeV (red), 100 MeV (blue) and 1 GeV (green).}
	\label{fig:intpl-tot}
\end{figure}

In Figure \ref{fig:intpl-tot} we show the total interaction rate for different values of the energy of the emitted photons: $E_\gamma = $ 1 MeV, 10 MeV, 100 MeV and 1 GeV.

\bibliography{references.bib}

\providecommand{\href}[2]{#2}\begingroup\raggedright\begin{thebibliography}{10}

\bibitem{Witten:1984rs}
E.~Witten, \emph{{Cosmic Separation of Phases}},
  \href{http://dx.doi.org/10.1103/PhysRevD.30.272}{\emph{Phys. Rev. D} {\bf 30}
  (1984) 272--285}.

\bibitem{Alcock:1985vc}
C.~Alcock and E.~Farhi, \emph{{The Evaporation of Strange Matter in the Early
  Universe}}, \href{http://dx.doi.org/10.1103/PhysRevD.32.1273}{\emph{Phys.
  Rev. D} {\bf 32} (1985) 1273}.

\bibitem{Alcock:1986hz}
C.~Alcock, E.~Farhi and A.~Olinto, \emph{{Strange stars}},
  \href{http://dx.doi.org/10.1086/164679}{\emph{Astrophys. J.} {\bf 310} (1986)
  261--272}.

\bibitem{Farhi:1984qu}
E.~Farhi and R.~L. Jaffe, \emph{{Strange Matter}},
  \href{http://dx.doi.org/10.1103/PhysRevD.30.2379}{\emph{Phys. Rev. D} {\bf
  30} (1984) 2379}.

\bibitem{LYNN1990186}
B.~W. Lynn, A.~E. Nelson and N.~Tetradis, \emph{Strange baryon matter},
  \href{http://dx.doi.org/https://doi.org/10.1016/0550-3213(90)90614-J}{\emph{Nuclear
  Physics B} {\bf 345} (1990) 186--209}.

\bibitem{Coleman:1985ki}
S.~R. Coleman, \emph{{Q Balls}},
  \href{http://dx.doi.org/10.1016/0550-3213(86)90520-1}{\emph{Nucl. Phys. B}
  {\bf 262} (1985) 263}.

\bibitem{Lee:1991ax}
T.~D. Lee and Y.~Pang, \emph{{Nontopological solitons}},
  \href{http://dx.doi.org/10.1016/0370-1573(92)90064-7}{\emph{Phys. Rept.} {\bf
  221} (1992) 251--350}.

\bibitem{Kusenko:1997si}
A.~Kusenko and M.~E. Shaposhnikov, \emph{{Supersymmetric Q balls as dark
  matter}}, \href{http://dx.doi.org/10.1016/S0370-2693(97)01375-0}{\emph{Phys.
  Lett. B} {\bf 418} (1998) 46--54},
  [\href{http://arxiv.org/abs/hep-ph/9709492}{{\tt hep-ph/9709492}}].

\bibitem{Kusenko:2004yw}
A.~Kusenko, L.~Loveridge and M.~Shaposhnikov, \emph{{Supersymmetric dark matter
  Q-balls and their interactions in matter}},
  \href{http://dx.doi.org/10.1103/PhysRevD.72.025015}{\emph{Phys. Rev. D} {\bf
  72} (2005) 025015}, [\href{http://arxiv.org/abs/hep-ph/0405044}{{\tt
  hep-ph/0405044}}].

\bibitem{Kusenko:1997vp}
A.~Kusenko, V.~Kuzmin, M.~E. Shaposhnikov and P.~G. Tinyakov,
  \emph{{Experimental signatures of supersymmetric dark matter Q balls}},
  \href{http://dx.doi.org/10.1103/PhysRevLett.80.3185}{\emph{Phys. Rev. Lett.}
  {\bf 80} (1998) 3185--3188}, [\href{http://arxiv.org/abs/hep-ph/9712212}{{\tt
  hep-ph/9712212}}].

\bibitem{Zhitnitsky:2006vt}
A.~Zhitnitsky, \emph{{Cold dark matter as compact composite objects}},
  \href{http://dx.doi.org/10.1103/PhysRevD.74.043515}{\emph{Phys. Rev. D} {\bf
  74} (2006) 043515}, [\href{http://arxiv.org/abs/astro-ph/0603064}{{\tt
  astro-ph/0603064}}].

\bibitem{Zhitnitsky:2006tu}
A.~Zhitnitsky, \emph{{The Width of the 511-KeV Line from the Bulge of the
  Galaxy}}, \href{http://dx.doi.org/10.1103/PhysRevD.76.103518}{\emph{Phys.
  Rev. D} {\bf 76} (2007) 103518},
  [\href{http://arxiv.org/abs/astro-ph/0607361}{{\tt astro-ph/0607361}}].

\bibitem{Cumberbatch:2006bj}
D.~T. Cumberbatch, J.~Silk and G.~D. Starkman, \emph{{Difficulties for Compact
  Composite Object Dark Matter}},
  \href{http://dx.doi.org/10.1103/PhysRevD.77.063522}{\emph{Phys. Rev. D} {\bf
  77} (2008) 063522}, [\href{http://arxiv.org/abs/astro-ph/0606429}{{\tt
  astro-ph/0606429}}].

\bibitem{Bai:2020jfm}
Y.~Bai, A.~J. Long and S.~Lu, \emph{{Tests of Dark MACHOs: Lensing, Accretion,
  and Glow}},
  \href{http://dx.doi.org/10.1088/1475-7516/2020/09/044}{\emph{JCAP} {\bf 09}
  (2020) 044}, [\href{http://arxiv.org/abs/2003.13182}{{\tt 2003.13182}}].

\bibitem{Cline:2013zca}
J.~M. Cline, Z.~Liu, G.~Moore and W.~Xue, \emph{{Composite strongly interacting
  dark matter}},
  \href{http://dx.doi.org/10.1103/PhysRevD.90.015023}{\emph{Phys. Rev. D} {\bf
  90} (2014) 015023}, [\href{http://arxiv.org/abs/1312.3325}{{\tt 1312.3325}}].

\bibitem{Carr:2016drx}
B.~Carr, F.~Kuhnel and M.~Sandstad, \emph{{Primordial Black Holes as Dark
  Matter}}, \href{http://dx.doi.org/10.1103/PhysRevD.94.083504}{\emph{Phys.
  Rev. D} {\bf 94} (2016) 083504}, [\href{http://arxiv.org/abs/1607.06077}{{\tt
  1607.06077}}].

\bibitem{doi:10.1146/annurev-nucl-050520-125911}
B.~Carr and F.~K\"uhnel, \emph{Primordial black holes as dark matter: Recent
  developments},
  \href{http://dx.doi.org/10.1146/annurev-nucl-050520-125911}{\emph{Annual
  Review of Nuclear and Particle Science} {\bf 70} (2020) 355--394}.

\bibitem{Abbott:2020khf}
{\scshape LIGO Scientific, Virgo} collaboration, R.~Abbott et~al.,
  \emph{{GW190814: Gravitational Waves from the Coalescence of a 23 Solar Mass
  Black Hole with a 2.6 Solar Mass Compact Object}},
  \href{http://dx.doi.org/10.3847/2041-8213/ab960f}{\emph{Astrophys. J. Lett.}
  {\bf 896} (2020) L44}, [\href{http://arxiv.org/abs/2006.12611}{{\tt
  2006.12611}}].

\bibitem{Bombaci:2020vgw}
I.~Bombaci, A.~Drago, D.~Logoteta, G.~Pagliara and I.~Vida\~na, \emph{{Was
  GW190814 a black hole -- strange quark star system?}},
  \href{http://dx.doi.org/10.1103/PhysRevLett.126.162702}{\emph{Phys. Rev.
  Lett.} {\bf 126} (2021) 162702}, [\href{http://arxiv.org/abs/2010.01509}{{\tt
  2010.01509}}].

\bibitem{Zhitnitsky:2002nr}
A.~R. Zhitnitsky, \emph{{Dark matter as dense color superconductor}},
  \href{http://dx.doi.org/10.1016/S0920-5632(03)02087-5}{\emph{Nucl. Phys. B
  Proc. Suppl.} {\bf 124} (2003) 99--102},
  [\href{http://arxiv.org/abs/astro-ph/0204218}{{\tt astro-ph/0204218}}].

\bibitem{Zhitnitsky:2004da}
A.~Zhitnitsky, \emph{{'Nonbaryonic' dark matter as baryonic colour
  superconductor}},
  \href{http://dx.doi.org/10.1088/0954-3899/30/1/061}{\emph{J. Phys. G} {\bf
  30} (2004) S513--S517}.

\bibitem{Liang:2016tqc}
X.~Liang and A.~Zhitnitsky, \emph{{Axion field and the quark nugget's formation
  at the QCD phase transition}},
  \href{http://dx.doi.org/10.1103/PhysRevD.94.083502}{\emph{Phys. Rev. D} {\bf
  94} (2016) 083502}, [\href{http://arxiv.org/abs/1606.00435}{{\tt
  1606.00435}}].

\bibitem{Bai:2018dxf}
Y.~Bai, A.~J. Long and S.~Lu, \emph{{Dark Quark Nuggets}},
  \href{http://dx.doi.org/10.1103/PhysRevD.99.055047}{\emph{Phys. Rev. D} {\bf
  99} (2019) 055047}, [\href{http://arxiv.org/abs/1810.04360}{{\tt
  1810.04360}}].

\bibitem{Bai:2018vik}
Y.~Bai and A.~J. Long, \emph{{Six Flavor Quark Matter}},
  \href{http://dx.doi.org/10.1007/JHEP06(2018)072}{\emph{JHEP} {\bf 06} (2018)
  072}, [\href{http://arxiv.org/abs/1804.10249}{{\tt 1804.10249}}].

\bibitem{Bai:2019ogh}
Y.~Bai and J.~Berger, \emph{{Nucleus Capture by Macroscopic Dark Matter}},
  \href{http://dx.doi.org/10.1007/JHEP05(2020)160}{\emph{JHEP} {\bf 05} (2020)
  160}, [\href{http://arxiv.org/abs/1912.02813}{{\tt 1912.02813}}].

\bibitem{Jacobs:2014yca}
D.~M. Jacobs, G.~D. Starkman and B.~W. Lynn, \emph{{Macro Dark Matter}},
  \href{http://dx.doi.org/10.1093/mnras/stv774}{\emph{Mon. Not. Roy. Astron.
  Soc.} {\bf 450} (2015) 3418--3430},
  [\href{http://arxiv.org/abs/1410.2236}{{\tt 1410.2236}}].

\bibitem{Sidhu:2019gwo}
J.~Singh~Sidhu, \emph{{Charge Constraints of Macroscopic Dark Matter}},
  \href{http://dx.doi.org/10.1103/PhysRevD.101.043526}{\emph{Phys. Rev. D} {\bf
  101} (2020) 043526}, [\href{http://arxiv.org/abs/1912.04732}{{\tt
  1912.04732}}].

\bibitem{SinghSidhu:2020cxw}
J.~Singh~Sidhu, R.~J. Scherrer and G.~Starkman, \emph{{Antimatter as
  Macroscopic Dark Matter}},
  \href{http://dx.doi.org/10.1016/j.physletb.2020.135574}{\emph{Phys. Lett. B}
  {\bf 807} (2020) 135574}, [\href{http://arxiv.org/abs/2006.01200}{{\tt
  2006.01200}}].

\bibitem{Kumar:2018rlf}
S.~Kumar, E.~Dimastrogiovanni, G.~D. Starkman, C.~Copi and B.~Lynn, \emph{{CMB
  Spectral Distortions from Cooling Macroscopic Dark Matter}},
  \href{http://dx.doi.org/10.1103/PhysRevD.99.023521}{\emph{Phys. Rev. D} {\bf
  99} (2019) 023521}, [\href{http://arxiv.org/abs/1804.08601}{{\tt
  1804.08601}}].

\bibitem{Sidhu:2019qoa}
J.~S. Sidhu, G.~Starkman and R.~Harvey, \emph{{Counter-top search for
  macroscopic dark matter}},
  \href{http://dx.doi.org/10.1103/PhysRevD.100.103015}{\emph{Phys. Rev. D} {\bf
  100} (2019) 103015}, [\href{http://arxiv.org/abs/1905.10025}{{\tt
  1905.10025}}].

\bibitem{Sidhu:2019fgg}
J.~S. Sidhu and G.~Starkman, \emph{{Macroscopic Dark Matter Constraints from
  Bolide Camera Networks}},
  \href{http://dx.doi.org/10.1103/PhysRevD.100.123008}{\emph{Phys. Rev. D} {\bf
  100} (2019) 123008}, [\href{http://arxiv.org/abs/1908.00557}{{\tt
  1908.00557}}].

\bibitem{Sidhu:2019oii}
J.~Singh~Sidhu, R.~J. Scherrer and G.~Starkman, \emph{{Death and serious injury
  from dark matter}},
  \href{http://dx.doi.org/10.1016/j.physletb.2020.135300}{\emph{Phys. Lett. B}
  {\bf 803} (2020) 135300}, [\href{http://arxiv.org/abs/1907.06674}{{\tt
  1907.06674}}].

\bibitem{Brandt:2016aco}
T.~D. Brandt, \emph{{Constraints on MACHO Dark Matter from Compact Stellar
  Systems in Ultra-Faint Dwarf Galaxies}},
  \href{http://dx.doi.org/10.3847/2041-8205/824/2/L31}{\emph{Astrophys. J.
  Lett.} {\bf 824} (2016) L31}, [\href{http://arxiv.org/abs/1605.03665}{{\tt
  1605.03665}}].

\bibitem{Sidhu:2019kpd}
J.~Singh~Sidhu and G.~D. Starkman, \emph{{Reconsidering astrophysical
  constraints on macroscopic dark matter}},
  \href{http://dx.doi.org/10.1103/PhysRevD.101.083503}{\emph{Phys. Rev. D} {\bf
  101} (2020) 083503}, [\href{http://arxiv.org/abs/1912.04053}{{\tt
  1912.04053}}].

\bibitem{Starkman:2020sbz}
N.~Starkman, J.~Sidhu, H.~Winch and G.~Starkman, \emph{{Straight Lightning as a
  Signature of Macroscopic Dark Matter}},
  \href{http://arxiv.org/abs/2006.16272}{{\tt 2006.16272}}.

\bibitem{Zumalacarregui:2017qqd}
M.~Zumalacarregui and U.~Seljak, \emph{{Limits on stellar-mass compact objects
  as dark matter from gravitational lensing of type Ia supernovae}},
  \href{http://dx.doi.org/10.1103/PhysRevLett.121.141101}{\emph{Phys. Rev.
  Lett.} {\bf 121} (2018) 141101}, [\href{http://arxiv.org/abs/1712.02240}{{\tt
  1712.02240}}].

\bibitem{Kuhnel:2017bvu}
F.~K\"uhnel, G.~D. Starkman and K.~Freese, \emph{{Primordial Black-Hole and
  Macroscopic Dark-Matter Constraints with LISA}},
  \href{http://arxiv.org/abs/1705.10361}{{\tt 1705.10361}}.

\bibitem{Richard:2019dic}
J.-M. Richard, \emph{{Antiproton physics}},
  \href{http://dx.doi.org/10.3389/fphy.2020.00006}{\emph{Front. in Phys.} {\bf
  8} (2020) 6}, [\href{http://arxiv.org/abs/1912.07385}{{\tt 1912.07385}}].

\bibitem{Consiglio:2017pot}
R.~Consiglio, P.~F. de~Salas, G.~Mangano, G.~Miele, S.~Pastor and O.~Pisanti,
  \emph{{PArthENoPE reloaded}},
  \href{http://dx.doi.org/10.1016/j.cpc.2018.06.022}{\emph{Comput. Phys.
  Commun.} {\bf 233} (2018) 237--242},
  [\href{http://arxiv.org/abs/1712.04378}{{\tt 1712.04378}}].

\bibitem{Cooke:2017cwo}
R.~J. Cooke, M.~Pettini and C.~C. Steidel, \emph{{One Percent Determination of
  the Primordial Deuterium Abundance}},
  \href{http://dx.doi.org/10.3847/1538-4357/aaab53}{\emph{Astrophys. J.} {\bf
  855} (2018) 102}, [\href{http://arxiv.org/abs/1710.11129}{{\tt 1710.11129}}].

\bibitem{Aver:2015iza}
E.~Aver, K.~A. Olive and E.~D. Skillman, \emph{{The effects of He I
  \ensuremath{\lambda}10830 on helium abundance determinations}},
  \href{http://dx.doi.org/10.1088/1475-7516/2015/07/011}{\emph{JCAP} {\bf 07}
  (2015) 011}, [\href{http://arxiv.org/abs/1503.08146}{{\tt 1503.08146}}].

\bibitem{Gariazzo:2021iiu}
S.~Gariazzo, P.~F. de~Salas, O.~Pisanti and R.~Consiglio, \emph{{PArthENoPE
  Revolutions}},  \href{http://arxiv.org/abs/2103.05027}{{\tt 2103.05027}}.

\bibitem{Aghanim:2018eyx}
{\scshape Planck} collaboration, N.~Aghanim et~al., \emph{{Planck 2018 results.
  VI. Cosmological parameters}},
  \href{http://dx.doi.org/10.1051/0004-6361/201833910}{\emph{Astron.
  Astrophys.} {\bf 641} (2020) A6},
  [\href{http://arxiv.org/abs/1807.06209}{{\tt 1807.06209}}].

\bibitem{Hu:1992dc}
W.~Hu and J.~Silk, \emph{{Thermalization and spectral distortions of the cosmic
  background radiation}},
  \href{http://dx.doi.org/10.1103/PhysRevD.48.485}{\emph{Phys. Rev. D} {\bf 48}
  (1993) 485--502}.

\bibitem{Chluba:2011hw}
J.~Chluba and R.~A. Sunyaev, \emph{{The evolution of CMB spectral distortions
  in the early Universe}},
  \href{http://dx.doi.org/10.1111/j.1365-2966.2011.19786.x}{\emph{Mon. Not.
  Roy. Astron. Soc.} {\bf 419} (2012) 1294--1314},
  [\href{http://arxiv.org/abs/1109.6552}{{\tt 1109.6552}}].

\bibitem{Chluba:2013vsa}
J.~Chluba, \emph{{Green's function of the cosmological thermalization
  problem}}, \href{http://dx.doi.org/10.1093/mnras/stt1025}{\emph{Mon. Not.
  Roy. Astron. Soc.} {\bf 434} (2013) 352},
  [\href{http://arxiv.org/abs/1304.6120}{{\tt 1304.6120}}].

\bibitem{Chluba:2016bvg}
J.~Chluba, \emph{{Which spectral distortions does $\Lambda$CDM actually
  predict?}}, \href{http://dx.doi.org/10.1093/mnras/stw945}{\emph{Mon. Not.
  Roy. Astron. Soc.} {\bf 460} (2016) 227--239},
  [\href{http://arxiv.org/abs/1603.02496}{{\tt 1603.02496}}].

\bibitem{Kogut:2019vqh}
A.~Kogut, M.~H. Abitbol, J.~Chluba, J.~Delabrouille, D.~Fixsen, J.~C. Hill
  et~al., \emph{{CMB Spectral Distortions: Status and Prospects}},
  \href{http://arxiv.org/abs/1907.13195}{{\tt 1907.13195}}.

\bibitem{Chluba:2019nxa}
J.~Chluba et~al., \emph{{New Horizons in Cosmology with Spectral Distortions of
  the Cosmic Microwave Background}},
  \href{http://arxiv.org/abs/1909.01593}{{\tt 1909.01593}}.

\bibitem{danese1982double}
L.~Danese and G.~De~Zotti, \emph{Double compton process and the spectrum of the
  microwave background}, {\emph{Astronomy and Astrophysics} {\bf 107} (1982)
  39--42}.

\bibitem{Mather:1993ij}
J.~C. Mather et~al., \emph{{Measurement of the Cosmic Microwave Background
  spectrum by the COBE FIRAS instrument}},
  \href{http://dx.doi.org/10.1086/173574}{\emph{Astrophys. J.} {\bf 420} (1994)
  439--444}.

\bibitem{Fixsen:1996nj}
D.~J. Fixsen, E.~S. Cheng, J.~M. Gales, J.~C. Mather, R.~A. Shafer and E.~L.
  Wright, \emph{{The Cosmic Microwave Background spectrum from the full COBE
  FIRAS data set}}, \href{http://dx.doi.org/10.1086/178173}{\emph{Astrophys.
  J.} {\bf 473} (1996) 576}, [\href{http://arxiv.org/abs/astro-ph/9605054}{{\tt
  astro-ph/9605054}}].

\bibitem{Kogut:2011xw}
A.~Kogut et~al., \emph{{The Primordial Inflation Explorer (PIXIE): A Nulling
  Polarimeter for Cosmic Microwave Background Observations}},
  \href{http://dx.doi.org/10.1088/1475-7516/2011/07/025}{\emph{JCAP} {\bf 07}
  (2011) 025}, [\href{http://arxiv.org/abs/1105.2044}{{\tt 1105.2044}}].

\bibitem{McDermott:2010pa}
S.~D. McDermott, H.-B. Yu and K.~M. Zurek, \emph{{Turning off the Lights: How
  Dark is Dark Matter?}},
  \href{http://dx.doi.org/10.1103/PhysRevD.83.063509}{\emph{Phys. Rev. D} {\bf
  83} (2011) 063509}, [\href{http://arxiv.org/abs/1011.2907}{{\tt 1011.2907}}].

\bibitem{Dubovsky:2003yn}
S.~L. Dubovsky, D.~S. Gorbunov and G.~I. Rubtsov, \emph{{Narrowing the window
  for millicharged particles by CMB anisotropy}},
  \href{http://dx.doi.org/10.1134/1.1675909}{\emph{JETP Lett.} {\bf 79} (2004)
  1--5}, [\href{http://arxiv.org/abs/hep-ph/0311189}{{\tt hep-ph/0311189}}].

\bibitem{Dolgov:2013una}
A.~D. Dolgov, S.~L. Dubovsky, G.~I. Rubtsov and I.~I. Tkachev,
  \emph{{Constraints on millicharged particles from Planck data}},
  \href{http://dx.doi.org/10.1103/PhysRevD.88.117701}{\emph{Phys. Rev. D} {\bf
  88} (2013) 117701}, [\href{http://arxiv.org/abs/1310.2376}{{\tt 1310.2376}}].

\bibitem{Dubovsky:2001tr}
S.~L. Dubovsky and D.~S. Gorbunov, \emph{{Small second acoustic peak from
  interacting cold dark matter?}},
  \href{http://dx.doi.org/10.1103/PhysRevD.64.123503}{\emph{Phys. Rev. D} {\bf
  64} (2001) 123503}, [\href{http://arxiv.org/abs/astro-ph/0103122}{{\tt
  astro-ph/0103122}}].

\bibitem{Melchiorri:2007sq}
A.~Melchiorri, A.~Polosa and A.~Strumia, \emph{{New bounds on millicharged
  particles from cosmology}},
  \href{http://dx.doi.org/10.1016/j.physletb.2007.05.042}{\emph{Phys. Lett. B}
  {\bf 650} (2007) 416--420}, [\href{http://arxiv.org/abs/hep-ph/0703144}{{\tt
  hep-ph/0703144}}].

\bibitem{Dvorkin:2013cea}
C.~Dvorkin, K.~Blum and M.~Kamionkowski, \emph{{Constraining Dark Matter-Baryon
  Scattering with Linear Cosmology}},
  \href{http://dx.doi.org/10.1103/PhysRevD.89.023519}{\emph{Phys. Rev. D} {\bf
  89} (2014) 023519}, [\href{http://arxiv.org/abs/1311.2937}{{\tt 1311.2937}}].

\bibitem{Tseliakhovich:2010bj}
D.~Tseliakhovich and C.~Hirata, \emph{{Relative velocity of dark matter and
  baryonic fluids and the formation of the first structures}},
  \href{http://dx.doi.org/10.1103/PhysRevD.82.083520}{\emph{Phys. Rev. D} {\bf
  82} (2010) 083520}, [\href{http://arxiv.org/abs/1005.2416}{{\tt 1005.2416}}].

\bibitem{Cyburt:2002uv}
R.~H. Cyburt, J.~R. Ellis, B.~D. Fields and K.~A. Olive, \emph{{Updated
  nucleosynthesis constraints on unstable relic particles}},
  \href{http://dx.doi.org/10.1103/PhysRevD.67.103521}{\emph{Phys. Rev. D} {\bf
  67} (2003) 103521}, [\href{http://arxiv.org/abs/astro-ph/0211258}{{\tt
  astro-ph/0211258}}].

\bibitem{Ishida:2014wqa}
H.~Ishida, M.~Kusakabe and H.~Okada, \emph{{Effects of long-lived 10 MeV-scale
  sterile neutrinos on primordial elemental abundances and the effective
  neutrino number}},
  \href{http://dx.doi.org/10.1103/PhysRevD.90.083519}{\emph{Phys. Rev. D} {\bf
  90} (2014) 083519}, [\href{http://arxiv.org/abs/1403.5995}{{\tt 1403.5995}}].

\bibitem{Poulin:2015woa}
V.~Poulin and P.~D. Serpico, \emph{{Loophole to the Universal Photon Spectrum
  in Electromagnetic Cascades and Application to the Cosmological Lithium
  Problem}},
  \href{http://dx.doi.org/10.1103/PhysRevLett.114.091101}{\emph{Phys. Rev.
  Lett.} {\bf 114} (2015) 091101}, [\href{http://arxiv.org/abs/1502.01250}{{\tt
  1502.01250}}].

\bibitem{Poulin:2015opa}
V.~Poulin and P.~D. Serpico, \emph{{Nonuniversal BBN bounds on
  electromagnetically decaying particles}},
  \href{http://dx.doi.org/10.1103/PhysRevD.91.103007}{\emph{Phys. Rev. D} {\bf
  91} (2015) 103007}, [\href{http://arxiv.org/abs/1503.04852}{{\tt
  1503.04852}}].

\bibitem{Salvati:2016jng}
L.~Salvati, L.~Pagano, M.~Lattanzi, M.~Gerbino and A.~Melchiorri,
  \emph{{Breaking Be: a sterile neutrino solution to the cosmological lithium
  problem}}, \href{http://dx.doi.org/10.1088/1475-7516/2016/08/022}{\emph{JCAP}
  {\bf 08} (2016) 022}, [\href{http://arxiv.org/abs/1606.06968}{{\tt
  1606.06968}}].

\bibitem{Zdziarski1989AbsorptionOX}
A.~Zdziarski and R.~Svensson, \emph{Absorption of x-rays and gamma rays at
  cosmological distances}, {\emph{The Astrophysical Journal} {\bf 344} (1989)
  551--566}.

\bibitem{Kawasaki:1994sc}
M.~Kawasaki and T.~Moroi, \emph{{Electromagnetic cascade in the early universe
  and its application to the big bang nucleosynthesis}},
  \href{http://dx.doi.org/10.1086/176324}{\emph{Astrophys. J.} {\bf 452} (1995)
  506}, [\href{http://arxiv.org/abs/astro-ph/9412055}{{\tt astro-ph/9412055}}].

\end{thebibliography}\endgroup

\end{document}